%% file: main.tex
\documentclass[journal]{IEEEtran}
\usepackage{cite}
\usepackage{hyperref}

\ifCLASSINFOpdf
  \usepackage[pdftex]{graphicx}
  \usepackage{algorithm}
  \graphicspath{{/figures/}{../pdf/}{../jpeg/}}
  \DeclareGraphicsExtensions{.pdf,.jpeg,.png}
\else
   \usepackage[dvips]{graphicx, caption, subcaption}
  \graphicspath{{figures/}{../pdf/}{../jpeg/}}
  \DeclareGraphicsExtensions{.eps,.png}
\fi
\usepackage[cmex10]{amsmath}
\usepackage[]{algorithm}
\usepackage{verbatim}
\usepackage{algpseudocode}
\usepackage{graphics}
\usepackage{caption}

\ifCLASSOPTIONcompsoc
  \usepackage[caption=false,font=normalsize,labelfont=sf,textfont=sf]{subfig}
\else
  \usepackage[caption=false,font=footnotesize]{subfig}
\fi
\usepackage{SIunits}

\usepackage{graphicx}

\begin{document}

%
\title{Localization of MEG and EEG Brain Signals by Alternating Projection}
%
%
%


\author{Amir Adler, Mati~Wax, Dimitrios Pantazis
\thanks{}
}
\maketitle

\begin{abstract}
We present a novel  solution to the problem of localization of MEG and EEG brain signals. The solution is sequential and iterative, and is based on minimizing the least-squares (LS) criterion by  the Alternating Projection (AP) algorithm, which is well known in the context of array signal processing. Unlike  existing  scanning solutions belonging to the beamformer and  multiple-signal classification (MUSIC) families, the algorithm has good performance in low signal-to-noise ratio (SNR) and can cope with closely spaced sources and any mixture of correlated sources. Results from simulated and experimental MEG data from a real phantom demonstrated robust performance across an extended SNR range, the entire inter-source correlation range, and across multiple sources, with consistently superior localization accuracy than popular scanning methods.
\end{abstract}

\begin{IEEEkeywords}
Brain signals, EEG, MEG, least-squares, alternating
 projections, iterative beamformer, MUSIC, iterative-sequential weighted-MUSIC,
iterative-sequential MUSIC, fully correlated sources, synchronous sources.
\end{IEEEkeywords}

\IEEEpeerreviewmaketitle
\section{Introduction}
\input{introduction.tex}

\section{Materials and Methods}
\subsection{Problem formulation}
\input{problem_formulation.tex}

\bibliographystyle{IEEEbib}
\bibliography{references}
\appendices  
\ifCLASSOPTIONcaptionsoff
  \newpage
\fi

\end{document}

%% file: introduction.tex
 \IEEEPARstart{B}{\lowercase{rain}} signals obtained by an array of magnetic or electric sensors using magnetoencephalography (MEG) or electroencephalography (EEG) offer the potential to investigate the complex spatiotemporal activity of the human brain. Localization of MEG and EEG sources has gained much interest in recent years since it can reveal the origins of neural signals in both the typical and atypical brain \cite{ilmoniemi2019brain}. 

Mathematically, the localization problem can be cast as an optimization problem of computing the location and moment parameters of the set of dipoles whose field best matches the MEG/EEG measurements in a least-squares (LS) sense \cite{mosher_multiple_1992}. In this paper we focus on \textit{dipole fitting} and \textit{scanning} methods, which solve for a small parsimonious set of dipoles and avoid the ill-posedness associated with imaging methods, such as minimum-norm \cite{hamalainen_magnetoencephalographytheory_1993}. Dipole fitting methods solve the optimization problem directly using techniques that include gradient descent, Nedler-Meade simplex algorithm, multistart, genetic algorithm, and simulated annealing \cite{huang_multi-start_1998,uutela_global_1998,khosla_spatio-temporal_1997,jiang_comparative_2003}. However, these techniques remain unpopular because they converge to a suboptimal local optimum or are too computationally expensive.

An alternative approach is scanning methods, which use adaptive spatial filters to search for optimal dipole positions throughout a discrete grid representing the source space \cite{darvas_mapping_2004}. Source locations are then determined as those for which a metric (localizer) exceeds a given threshold. While these approaches do not lead to true least squares solutions, they can be used to initialize a local least squares search. The most common scanning methods are beamformers \cite{LCMV,VerbaRobinson} and MUSIC \cite{mosher_multiple_1992}, both widely used for bioelectromagnetic source localization, but they assume uncorrelated sources. When correlations are significant, they result in partial or complete cancellation of correlated (also referred to as synchronous) sources) and distort the estimated time courses. Several multi-source extensions have been proposed for synchronous sources \cite{CohBF, CorrBF, NullBF, DCBF, EDCBF, MultiLCMV, MultiBeamformers, POP-MUSIC, WedgeMUSIC}, however they require some a-priori information on the location of the synchronous sources, are limited to the localization of pairs of synchronous sources, or are limited in their performance.

One important division of the scanning methods is whether they are \textit{non-recursive} or \textit{recursive}. The original Beamformer \cite{LCMV,VerbaRobinson} and MUSIC \cite{mosher_multiple_1992} methods are non-recursive and require the identification of the largest local maxima in the localizer function to find multiple dipoles. Some multi-source variants are also non-recursive (e.g. \cite{hui_identifying_2010, CorrBF, DCBF, EDCBF}), and as a result they use brute-force optimization, assume that the approximate locations of the neuronal sources have been identified a priori, or still require the identification of the largest local maxima in the localizer function. To overcome these limitations, non-recursive methods have recursive counterparts, such as RAP MUSIC \cite{Mosher1999}, TRAP MUSIC \cite{makela_truncated_2018}, Recursive Double-Scanning MUSIC \cite{makela_locating_2017}, and RAP Beamformer \cite{ilmoniemi2019brain}. The idea behind the recursive execution is that one finds the source locations iteratively at each step, projecting out the topographies of the previously found dipoles before forming the localizer for the current step \cite{Mosher1999,ilmoniemi2019brain}. In this way, one replaces the task of finding several local maxima with the easier task of finding the global maximum of the localizer at each iteration step. While recursive methods generally perform better than their non-recursive counterparts, they still suffer from several limitations, including limited performance, the need for high signal-to-noise ratio (SNR), non-linear optimization of source orientation angles and source amplitudes, or inaccurate estimation as correlation values increase.

 Here we present a novel solution to the source localization problem which does not have these limitations.  Our starting point is the LS estimation criterion for the dipole fitting problem. As is well known, this criterion yields a multidimensional nonlinear and nonconvex minimization problem, making it very challenging to avoid being trapped in undesirable local minima \cite{SPMBrainMap}. To overcome this challenge, we propose the use of the Alternating Projection (AP) algorithm \cite{AP}, which is well known in sensor array signal processing. The AP algorithm transforms the multidimensional problem to an iterative process involving only one-dimensional maximization problems, which are computationally much simpler. Moreover, the algorithm has a very effective initialization scheme which is the key to its good global convergence. 
 
 Since the AP algorithm minimizes the LS criterion, which coincides with the maximum likelihood (ML) criterion in the case of white and Gaussian noise, it affords the following properties: (i) good performance in low SNR and in low number of samples, not failing even in the case of a single sample, (ii) good resolution between closely spaced sources, and (iii) ability to cope with any mixture of correlated sources, including perfectly synchronous sources. 
 
 We note that the AP algorithm is conceptually close to the recursive scanning methods, in that it estimates the source locations iteratively at each step, projecting out the topographies of the previously found dipoles before forming the localizer for the current step. A fundamental difference between the AP algorithm and the recursive scanning methods is that the AP algorithm iterates through all sources multiple times until convergence while the recursive methods end when the last source is found. As such, beyond the AP algorithm which performs a LS estimation, there can be AP extensions of the recursive scanning methods. For example, sequential MUSIC (S-MUSIC) \cite{S-MUSIC} naturally extends to AP MUSIC as we describe in the method section. 

In this paper, we derive the AP source localization method and relate the analytical results against the standard beamformer and MUSIC scanning methods. We then compare the performance of AP against popular MUSIC and beamformer variants in simulations that vary SNR and inter-source correlation levels, and for different numbers of dipole sources. Finally, we validate the accuracy of the AP method and characterize its localization properties with experimental MEG phantom data.

%% file: problem_formulation.tex
In this section we briefly review the notations used to describe measurement data, forward matrix, and sources, and formulate the problem of estimating current dipoles. Consider an array of $M$ MEG or EEG sensors that measures data from a finite number $Q$ of equivalent current dipole (ECD) sources emitting signals $\{s_q(t)\}^{Q}_{q=1}$ at locations $\{\mathbf p_q\}^{Q}_{q=1}$. Under these assumptions, the $M\times 1$ vector of the received signals by the array is given by:
\begin{equation}
\mathbf y(t) = \sum_{q=1}^{Q} \mathbf l(\mathbf p_q)  s_{q}(t) +\mathbf n(t),
\label{eq:snapshot1}
\end{equation}
where  $\mathbf l(\mathbf p_q)$ is the topography of the dipole at location $\mathbf p_q$  and $\mathbf n(t)$  is the additive noise. The topography $\mathbf l(\mathbf p_q)$, is given by: 
\begin{equation}
\label{topography_model}
\mathbf l(\mathbf p_q) = \mathbf L(\mathbf p_q)  \mathbf{q},
\end{equation}
where $\mathbf L (\mathbf p_q)$ is the $M\times 3$ forward matrix at location $\mathbf p_q$ and $\mathbf q$ is the $3\times 1$ vector of the orientation of the ECD source. Depending on the problem, the orientation $\mathbf q$ may be  known, referred to as \textit{fixed-oriented} dipole, or it may be  unknown, referred to as \textit{ freely-oriented} dipole.

Assuming that the array is sampled  $N$ times at $t_1,...,t_N$, the matrix $\mathbf Y$ of the sampled signals can be expressed as: 
\begin{equation}
\label{basic_equation}
\mathbf Y =\mathbf A(\mathbf P) \mathbf S +\mathbf N, 
\end{equation} 
where  $\mathbf Y$ is the $M\times N$ matrix of the received signals:
\begin{equation}
\mathbf Y= [\mathbf y(t_1), ..., \mathbf y (t_N)],
\label{eq:snapshot4}
\end{equation}
$\mathbf A(\mathbf P)$ is the $M\times Q$ mixing matrix of the topography vectors at the $Q$ locations $\mathbf P=[\mathbf p_1,...,\mathbf p_Q]$:
\begin{equation}
\mathbf A(\mathbf P)=[\mathbf l(\mathbf p_1), ..., \mathbf l(\mathbf p_Q)],
\label{eq:snapshot5}
\end{equation}
$\mathbf S$ is the $Q\times N$ matrix of the sources:
\begin{equation}
\mathbf S= [\mathbf s(t_1), ..., \mathbf s (t_N)],
\label{eq:snapshot6}
\end{equation}
with $\mathbf s(t)=[s_1(t),..., s_Q(t)]^{T}$, and $\mathbf N$ is the $M\times N$ matrix of noise:
\begin{equation}
\mathbf N =[\mathbf n(t_1),..., \mathbf n(t_N)].
\label{eq:snapshot8}
\end{equation}

We further make the following assumptions regarding  the emitted signals and the propagation model:

A1:  The number of sources $Q$ is \textit{known} and obeys $Q<M$.

A2:  The emitted signals are  \textit{unknown} and  \textit{arbitrarily correlated}, including  the case that a subset of the sources or all of them are \textit{synchronous}. 

A3: The forward matrix $\mathbf L(\mathbf p)$ is \textit{ known}   for every location $\mathbf p$ (computed by the \textit{forward model}). 

A4: Every  $Q$ topography vectors $\{\mathbf l(\mathbf p_q)\}_{q=1}^Q$ are linearly independent, i.e., $\mathrm{rank}\mathbf A(\mathbf P)=Q$.

\hfill
\hfill

We can now state the problem of localization of brain signals as follows: \textit{Given the received data $\mathbf Y$, estimate the $Q$ locations of the sources $\{\mathbf p_q\}_{q=1}^{Q}$}.

\subsection{The Alternating Projection solution } 
We first solve the problem for the fixed-oriented dipoles and then extend it to freely-oriented dipoles. The least-squares estimation criterion, from \eqref{basic_equation},    is  given by 

\begin{equation}
\label{LS_criterion}
\{\hat{\mathbf P},\hat{\mathbf S}\}_\text{ls} = \arg\min_{\mathbf P, \mathbf S}    \|\ \mathbf Y-\mathbf A(\mathbf P)  \mathbf S \|^{2}_{F}.
\end{equation}
where $F$ denotes the Frobenius norm. To solve this minimization problem, we first eliminate the unknown signal matrix $\mathbf S$ by expressing it in terms of  $\mathbf P$. To this end, we equate  to zero the derivative of \eqref{LS_criterion} with respect to $\mathbf S$, and solve for $\mathbf S$:
\begin{equation}
\label{s_hat}
\mathbf{\hat  S} =  \mathbf (\mathbf A(\mathbf P)^{T}\mathbf A(\mathbf P) )^{-1}\mathbf A(\mathbf P)^{T}  \mathbf Y.
\end{equation}
Defining $\mathbf{\Pi_{\mathbf A(\mathbf P)}} $  the projection matrix on the column span of $\mathbf A( \mathbf P)$ 
\begin{equation}
\label{Projection_A}
\mathbf \Pi_{\mathbf A(\mathbf P)}=\mathbf A(\mathbf P)(\mathbf A(\mathbf P)^T\mathbf A(\mathbf P))^{-1}\mathbf A(\mathbf P)^T,
\end{equation}
equations \eqref{LS_criterion}, \eqref{s_hat}, and \eqref{Projection_A} now yield:
\begin{eqnarray}
{\mathbf{\hat  P}} & = & \arg\min_{\mathbf P}  \|\ (\mathbf I -  \mathbf{\Pi}_{\mathbf A(\mathbf P)}) \mathbf Y \|^{2}_{F} \\
& = & \arg\max_{\mathbf P}  \|  \mathbf {\Pi}_{\mathbf A(\mathbf P)} \mathbf Y \|^{2}_{F} \\
& = & \arg\max_{\mathbf P }\mathrm{tr} (\mathbf{\Pi}_{\mathbf A(\mathbf P)} \mathbf C  \mathbf{\Pi}_{\mathbf A(\mathbf P)}) \\
\label{LS_cost}
& = & \arg\max_{\mathbf P} \mathrm{tr} (\mathbf{\Pi}_{\mathbf A(\mathbf P)} \mathbf C) 
\end{eqnarray}
where $\rm tr( \;)$ denotes the trace operator, subscript $F$ denotes the Frobenius norm, $\mathbf C$ is the data covariance matrix $\mathbf C= \mathbf Y\mathbf Y^T$, and $\mathbf I$ denotes the identity matrix.

This is a nonlinear and nonconvex $Q$-dimensional maximization problem. The AP algorithm \cite{AP} solves this problem by transforming it to a sequential and iterative process involving only 1-dimensional maximization problems. The transformation is based on the projection-matrix decomposition formula. Let $\mathbf B$ and $\mathbf D$ be two matrices with the same number of rows, and let $\mathbf \Pi_{[\mathbf B,\mathbf D]}$ denote the projection-matrix onto the column span of the augmented matrix $[\mathbf B, \mathbf D ]$. Then:
\begin{equation}
\label{Proj_decomp}
\mathbf \Pi_{[\mathbf B,\mathbf D]}= \mathbf \Pi_{\mathbf B}+ \mathbf \Pi^{\perp}_{\mathbf B} \mathbf \Pi_{\mathbf D}, 
\end{equation}   
where $\mathbf \Pi^{\perp}_{\mathbf B}$ is the projection to the orthogonal complement of $\mathbf B$, given by $\mathbf \Pi^{\perp}_{\mathbf B} = (\mathbf I -\mathbf \Pi_{\mathbf B})$. 
 
 The AP algorithm exploits this  decomposition to transform the multidimensional maximization \eqref{LS_cost} into a sequential and iterative  process involving only maximization over a single parameter at a time, with all the other parameters held fixed at their pre-estimated values. More specifically, let $j+1$ denote the current iteration number, and let $q$ denote the current source to be estimated (q is sequenced from 1 to $Q$ in every iteration). The other sources are held fixed at their pre-estimated values: $\{\hat{ \mathbf p}_i^{(j+1)}\}_{i=1}^{q-1}$, which have been pre-estimated in the current iteration, and $\{\hat{ \mathbf p}_i^{(j)}\}_{i=q+1}^{Q}$, which have been pre-estimated in the previous iteration. With this notation,
 let $\mathbf A(\mathbf { \hat {P}}_{(q)}^{(j)})$ denote the $M\times (Q-1)$ matrix of the topographies corresponding to these values (note that  the $q$-th topography is excluded), given by:
\begin{equation}
\label{eq:R_i_test}
\mathbf A(\mathbf { \hat {P}}_{(q)}^{(j)})=[\mathbf l( \hat {\mathbf p}_{1}^{(j+1)}),...,\mathbf l(\hat {\mathbf p}_{q-1}^{(j+1)}),\mathbf l(\hat {\mathbf p}_{q+1}^{(j)}),...,\mathbf l(\hat {\mathbf p}_{Q}^{(j)})].
\end{equation} 
By the projection matrix decomposition \eqref{Proj_decomp}, we have:
\begin{equation}
\label{Proj_decomp_A}
\mathbf{\Pi}_{[\mathbf A(\mathbf { \hat {P}}_{(q)}^{(j)}),\mathbf l(\mathbf p_q)] }= \mathbf{\Pi}_{\mathbf A(\mathbf { \hat {P}}_{(q)}^{(j)})}+ \mathbf{\Pi}_{\mathbf A(\mathbf { \hat {P}}_{(q)}^{(j)})}^{\perp} \mathbf \Pi_{\mathbf l(\mathbf p_q)}.
\end{equation}
Substituting \eqref{Proj_decomp_A}  into \eqref{LS_cost}, and ignoring the contribution of the first term since it is not a function of $\mathbf p_q$, we get:
\begin{equation}
\label{ap_intermediate_solution}
\mathbf  {\mathbf{\hat p}}_{q}^{(j+1)} =\arg\max_{\mathbf p_q}\mathrm{tr} (\mathbf{\Pi}_{\mathbf A(\mathbf { \hat {P}}_{(q)}^{(j)})}^{\perp} \mathbf \Pi_{\mathbf l(\mathbf p_q)}\mathbf C).
\end{equation}
To ease notation, we define:
\begin{equation}
\mathbf Q_{(q)}^{(j)}= \mathbf{\Pi}_{\mathbf A(\mathbf { \hat {P}}_{(q)}^{(j)})}^{\perp}, 
\end{equation}
with $\mathbf Q_{(q)}^{(j)}$ denoting the projection matrix that projects out all but the $q$-th source at iteration $j$. Then \eqref{ap_intermediate_solution}  can be rewritten, using the properties of the projection and  trace operators, as:
\begin{equation}
\label{Iterative_solution}
\mathbf  {\mathbf{\hat p}}_{q}^{(j+1)} =\arg\max_{\mathbf p_q}\frac{\mathbf l^T(\mathbf p_q)\mathbf Q_{(q)}^{(j)}\mathbf C \mathbf Q_{(q)}^{(j)}\mathbf l(\mathbf p_q)}{{\mathbf l^T(\mathbf p_q) \mathbf Q_{(q)}^{(j)}\mathbf l(\mathbf p_q)}}.
\end{equation}

The initialization of the algorithm, which is critical for its good global convergence, is very straightforward. First we solve \eqref{LS_cost} for a single source, yielding
\begin{equation}
\label{eq:p_q3}
  \mathbf{\hat p}_{1}^{(0)} =\arg\max_{\mathbf p_1}\frac{\mathbf l^T(\mathbf p_1)\mathbf C \mathbf l(\mathbf p_1)}{{\mathbf l^T(\mathbf p_1) \mathbf l(\mathbf p_1)}}.
\end{equation}
Then, we  add one source at a time and solve for the $q$-th source, $q=2,...,Q$, yielding
\begin{equation}
\label{eq:p_q4}
\mathbf  {\mathbf{\hat p}}_{q}^{(0)} =\arg\max_{\mathbf p_q}\frac{\mathbf l^T(\mathbf p_q)\mathbf Q^{(0)}_{(q)}\mathbf C \mathbf Q^{(0)}_{(q)}\mathbf l(\mathbf p_q)}{{\mathbf l^T(\mathbf p_q) \mathbf Q^{(0)}_{(q)}\mathbf l(\mathbf p_q)}},
\end{equation}
where $\mathbf Q^{(0)}_{(q)}$ is the projection matrix that projects out the previously estimated $q-1$ sources: 
\begin{equation}
\mathbf Q^{(0)}_{(q)}= (\mathbf I -\mathbf{\Pi}_{\mathbf A(\mathbf { \hat {P}}^{(0)}_{(q)})}), 
\end{equation}
with $\mathbf A(\mathbf { \hat {P}}^{(0)}_{(q)})$ being the $M\times(q-1)$ matrix given by
\begin{equation}
\label{eq:R_i_test7}
\mathbf A(\mathbf { \hat {P}}_{(q)}^{(0)})=[\mathbf l( \hat {\mathbf p}_{1}^{(0)}),...,\mathbf l(\hat {\mathbf p}_{q-1}^{(0)})].
\end{equation} 
Once the initial location of the $Q$-th source has been estimated, subsequent iterations of the algorithm, described by \eqref{Iterative_solution}, begin. The iterations continue till the localization refinement from one iteration to the next is below a pre-defined threshold. 

Note that the algorithm climbs the peak of \eqref{LS_cost} along lines parallel to  $\mathbf p_1,...\mathbf p_Q$, with the climb rate depending on the structure of the cost function in the proximity of the peak. Since a maximization is performed at every step, the value of the maximized function cannot decrease. As a result, the algorithm is bound to converge to a local maximum which may not necessarily be the global one. Yet, in practice the above described initialization procedure yields good global convergence.

In the case of \textit{freely-oriented} dipoles, it follows from \eqref{topography_model} that the solution \eqref{Iterative_solution}  becomes:
\begin{equation}
\label{eq:p_q6}
\mathbf  {\mathbf{\hat p}}_{q}^{(j+1)} =\arg\max_{\mathbf p_q, \mathbf q}\frac{\mathbf  q^T \mathbf L^T(\mathbf p_q)\mathbf Q_{(q)}^{(j)}\mathbf C \mathbf Q_{(q)}^{(j)}\mathbf L(\mathbf p_q)\mathbf q}{{\mathbf q^T\mathbf L^T(\mathbf p_q) \mathbf Q_{(q)}^{(j)}\mathbf L(\mathbf p_q)\mathbf q}},
\end{equation}
whose solution is given by
\begin{equation}
  \hat{\mathbf q}^{(j+1)}
= \arg\max_{\mathbf p_q} 
 \mathbf v_1(\mathbf F_{(q)}^{(j)},\mathbf G_{(q)}^{(j)}),
\end{equation}
and
\begin{equation}
  \hat{\mathbf p}^{(j+1)}_q
=  
 \arg\max_{\mathbf p_q}\lambda_1(\mathbf F_{(q)}^{(j)},\mathbf G_{(q)}^{(j)}),
\end{equation}
where 
\begin{equation}
 \mathbf F_{(q)}^{(j)}= \mathbf L^T(\mathbf p_q)\mathbf Q_{(q)}^{(j)}\mathbf C \mathbf Q_{(q)}^{(j)}\mathbf L(\mathbf p_q),  
\end{equation}
and
\begin{equation}
  \mathbf G_{(q)}^{(j)}=\mathbf L^T(\mathbf p_q)\mathbf Q_{(q)}^{(j)} \mathbf Q_{(q)}^{(j)}\mathbf L(\mathbf p_q),  
\end{equation}
with $\mathbf v_1(\mathbf F,\mathbf G)$ and $\lambda_1(\mathbf F,\mathbf G)$ denoting  the generalized eigenvector and generalized eigenvalue, respectively, corresponding to the maximum generalized eigenvalue of the matrix pencil $(\mathbf F,\mathbf G)$.

\subsection{Relationship of Alternating Projection to other localization methods}

It is interesting to compare the above derived  AP  algorithms to the seemingly similar recursively applied and projected (RAP) beamformer, which is shown in \cite{ilmoniemi2019brain} to be mathematically equivalent to the multiple constrained minimum variance (MCMV) \cite{MultiBeamformers}. To this end,  we write down the  RAP beamformer for a fixed-oriented dipole as
\begin{equation}
\label{RAP_beamformer}
\mathbf  {\mathbf{\hat p}}_{q} =\arg\max_{\mathbf p_q}\frac{\mathbf l^T(\mathbf p_q) \mathbf Q^{(0)}_{(q)}\mathbf l(\mathbf p_q)}{{\mathbf l^T(\mathbf p_q) (\mathbf Q^{(0)}_{(q)}\mathbf C\mathbf Q^{(0)}_{(q)})^{\dagger}\mathbf l(\mathbf p_q)}},
\end{equation}
where $\dagger$ denotes the pseudo-inverse. 
Comparing it to the AP solution, the nominator  is identical to the denominator of \eqref{Iterative_solution}, while  the denominator, excluding the pseudoinverse, is identical to the nominator of \eqref{Iterative_solution}. Note that the pseudoinverse "compensates" for this  role switching of the nominator and the denominator, making the two algorithms  seemingly similar.  Computationally, \eqref{Iterative_solution} is much simpler than \eqref{RAP_beamformer} since no psuedoinverse is required. Moreover, it works even  for the case of a single sample and for the case of synchronous  sources, i.e., when $\mathbf C$ is rank 1, in which case the RAP beamformer, as all the LCMV-based beamformers, fails.  Another difference between the algorithms is their iterative nature. While the RAP beamformer \eqref{RAP_beamformer} is computed  only $Q$ times, for $q=1,...,Q$,  the AP algorithm \eqref{Iterative_solution}  is computed iteratively till convergence, thus enabling improved performance. A similar multi-iteration version of the MCMV solution was presented in \cite{MultiLCMV2}.

Since the number of sources $Q$ is assumed known, we can  replace the covariance matrix $\mathbf C$ in \eqref{Iterative_solution} by its signal-subspace approximation given by 
$\mathbf U_s \mathbf \Lambda_s \mathbf U_s^T$,  where $\mathbf \Lambda_s$ is the matrix of the $Q$ largest eigenvalues of $\mathbf C$, and $\mathbf U_s$ is the matrix of the corresponding eigenvectors. We get:
\begin{equation}
\label{eq:p_q13}
\mathbf  {\mathbf{\hat p}}_{q}^{(j+1)} =\arg\max_{\mathbf p_q}\frac{\mathbf l^T(\mathbf p_q)\mathbf Q_{(q)}^{(j)} \mathbf U_s \mathbf \Lambda_s \mathbf U_s^T \mathbf Q_{(q)}^{(j)}\mathbf l(\mathbf p_q)}{{\mathbf l^T(\mathbf p_q) \mathbf Q_{(q)}^{(j)}\mathbf l(\mathbf p_q)}},
\end{equation}
This algorithm can be regarded as the iterative and sequential version of the weighted-MUSIC algorithm \cite{WeightedMUSIC}, and we refer to it as AP-wMUSIC. 

If we omit the weighting matrix $\mathbf \Lambda_s$ from this expression we get:
\begin{equation}
\label{AP_MUSIC}
\mathbf  {\mathbf{\hat p}}_{q}^{(j+1)} =\arg\max_{\mathbf p_q}\frac{\mathbf l^T(\mathbf p_q)\mathbf Q_{(q)}^{(j)} \mathbf U_s  \mathbf U_s^T \mathbf Q_{(q)}^{(j)}\mathbf l(\mathbf p_q)}{{\mathbf l^T(\mathbf p_q) \mathbf Q_{(q)}^{(j)}\mathbf l(\mathbf p_q)}},
\end{equation}
which is the same form as the sequential MUSIC (S-MUSIC) \cite{S-MUSIC}, with the difference that \eqref{AP_MUSIC} is computed iteratively till convergence while the S-MUSIC is computed only  $Q$ times for $ q=1,...Q$, with no iterations.  We refer to \eqref{AP_MUSIC} as \textit{AP-MUSIC}.
We should point out that an algorithm similar to \eqref{AP_MUSIC}, referred to as self-consistent MUSIC has been introduced in \cite{SelfConsistent}. Yet, being based on RAP-MUSIC \cite{Mosher1999}, that algorithm is computationally more complex.

When all sources are known to be \textit{synchronous} \cite{31267}, one can improve the proposed estimators.  Indeed, synchronous sources yield a signal subspace of rank 1, and hence $\mathbf U_S$ must be replaced by $\mathbf u_1$ in both the AP-wMUSIC and in AP-MUSIC algorithms. This results in an identical solution for both:
\begin{equation}
\label{Synchronous}
\mathbf  {\mathbf{\hat p}}_{q}^{(j+1)} =\arg\max_{\mathbf p_q}\frac{\mathbf l^T(\mathbf p_q)\mathbf Q_{(q)}^{(j)} \mathbf u_1 \mathbf u_1^T \mathbf Q_{(q)}^{(j)}\mathbf l(\mathbf p_q)}{{\mathbf l^T(\mathbf p_q) \mathbf Q_{(q)}^{(j)}\mathbf l(\mathbf p_q)}}.
\end{equation}

\begin{figure}
    \centering
    \captionsetup{position=top}
    \subfloat[Simulation geometry]{\includegraphics[trim=0.5cm 0.0cm 0.75cm 0.0cm, clip,scale=0.2]{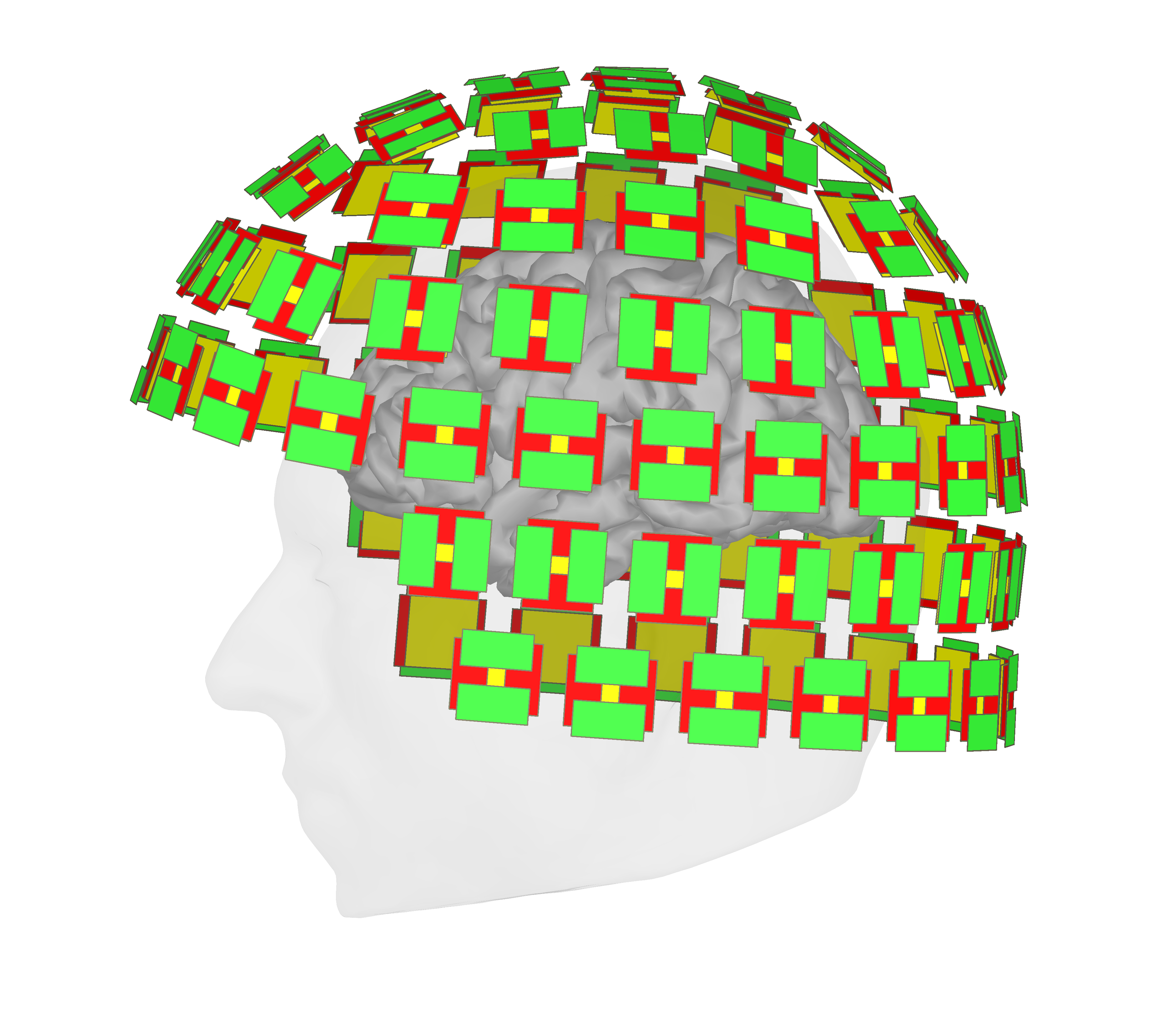}}
    \\
    \vspace{-.2cm}
    \subfloat[Phantom geometry]{
    \includegraphics[trim=0.5cm 0.0cm 0.75cm 0.0cm, clip,scale=0.40]{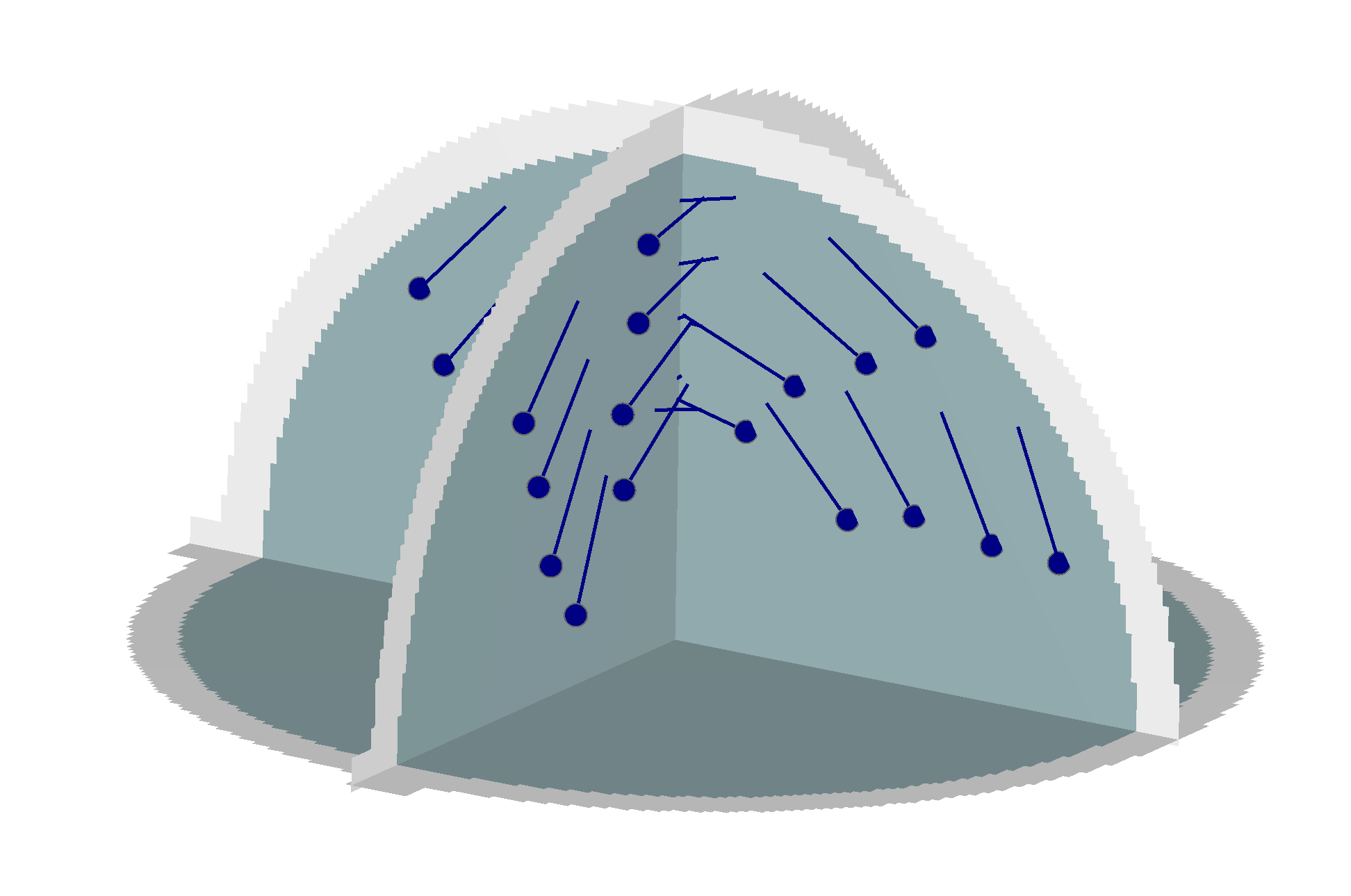}}
    \vspace{-0.2cm}
    \caption{Geometry of simulation and phantom experiments. (a) Simulations used the anatomy of an adult human subject and a whole-head MEG sensor array from an Elekta Triux device. (b) Real phantom experiments activated an array of 32 artificial dipoles in the depicted locations, and signals were measured with a whole-head MEG sensor array from an Elekta Neuromag device.}
    \label{fig:geometry}
\end{figure}

\begin{figure*}[ht!]
    \centering
    \captionsetup{position=top}
    \subfloat[Two sources with inter-source correlation $\rho=0$]{\includegraphics[trim=0.5cm 0.0cm 0.75cm 0.0cm, clip,scale=0.55]{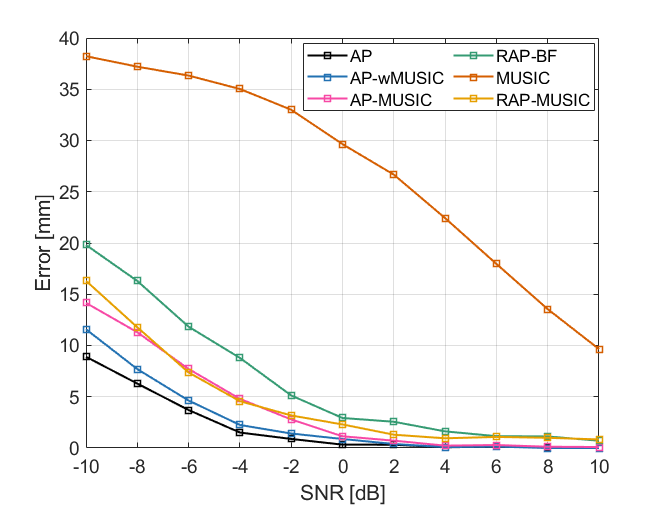}
    \label{subfig1}}
    \subfloat[Two sources with inter-source correlation $\rho=0.5$]{\includegraphics[trim=0.5cm 0.0cm 0.75cm 0.0cm, clip,scale=0.55]{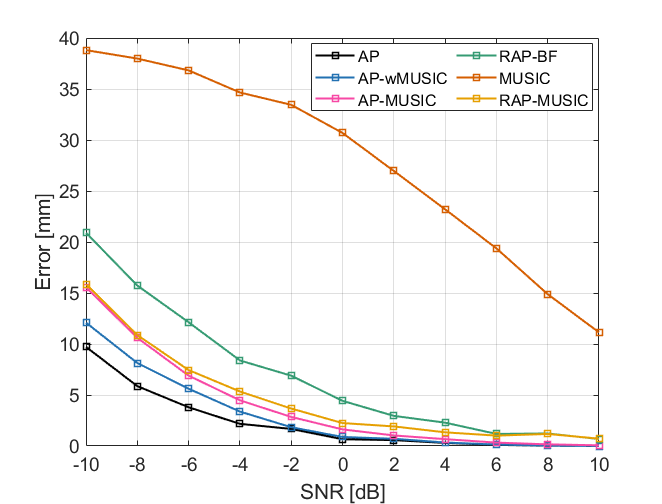}}
    \caption{Source localization error for an extended range of signal-to-noise ratio values. Monte Carlo simulations were conducted for the case of two sources with inter-source correlation 0 (a) and 0.5 (b).}
    \label{fig:snr}
\end{figure*}

\begin{figure*}[ht!]
    \centering
    \captionsetup{position=top}
    \subfloat[Two sources with SNR -10 dB]{\includegraphics[trim=0.5cm 0.0cm 0.75cm 0.0cm, clip,scale=0.55]{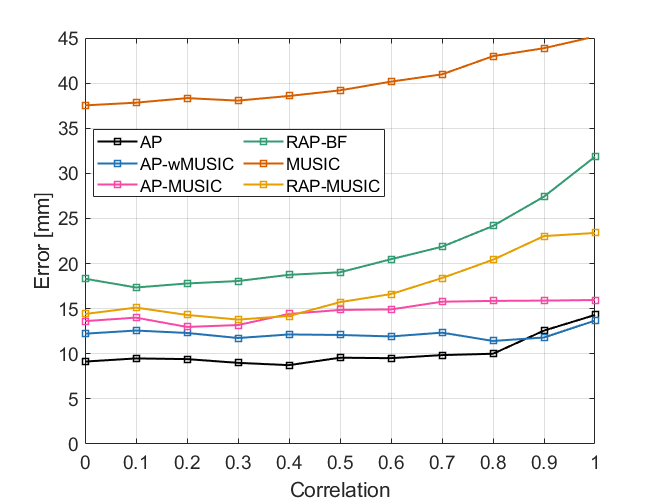}}
    \subfloat[Two sources with SNR 0 dB]{    \includegraphics[trim=0.5cm 0cm 0.75cm 0cm, clip,scale=0.55]{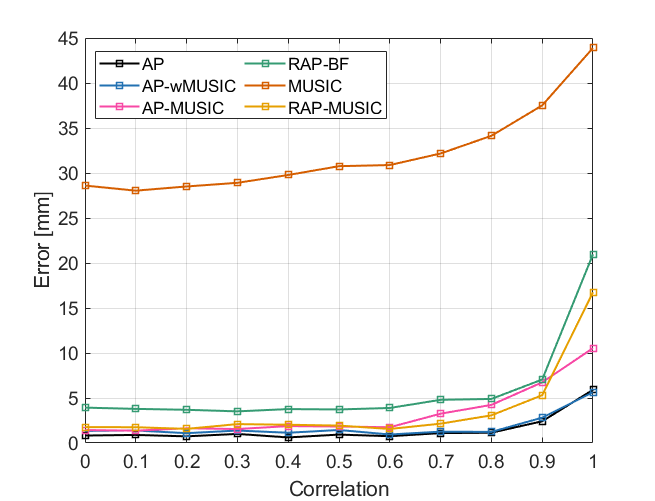}}
    \caption{Source localization error across the entire inter-source correlation range. Monte Carlo simulations were conducted for the case of two sources with signal-to-noise ratio -10 dB (a) and 0 dB (b).}
    \label{fig:correlation}
\end{figure*}

\subsection{Performance evaluation with simulations}

We assessed the performance of the AP localization method against five alternative source localization methods: MUSIC, RAP-MUSIC, RAP-beamformer, AP-MUSIC, and AP-wMUSIC. The last two are variants of the original MUSIC algorithm using the AP algorithm to perform iterations until convergence, and were introduced in the previous section. We compared the performance of all methods for different source configurations of varying levels of SNR and inter-source correlation, defined by:
\begin{equation}
    \rho=\frac{\textbf{a}^T\textbf{b}}{\|\textbf{a}\|_2\|\textbf{b}\|_2},
\end{equation}
where $\textbf{a}$ and $\textbf{b}$ are the corresponding sources' waveform vectors.

The geometry of the sensor array was based on the whole-head Elekta Triux MEG system (306-channel probe unit with 204 planar gradiometer sensors and 102 magnetometer sensors) (Figure \ref{fig:geometry}a). The geometry of the MEG source space was modeled with the cortical manifold extracted from a real adult human subject's MR data using Freesurfer \cite{fischl_2004}. Sources were restricted to approximately 15,000 grid points over the cortex. The lead field matrix was estimated using BrainStorm \cite{Brainstorm2011} based on an overlapping spheres head model \cite{Huang1999}. 
Gaussian white noise was generated and added to the MEG sensors to model instrumentation noise at specified SNR levels. The SNR was defined as the ratio of the Frobenius norm of the signal-magnetic-field spatiotemporal matrix to that of the noise matrix for each trial as in \cite{sekihara2001}.

We first considered cases where the number of dipole sources and inter-source correlation were fixed, and evaluated localization accuracy for a range of SNR values. Subsequently, we fixed the SNR level and varied inter-source correlation to assess robustness in localization even for highly synchronous sources. Last, we increased the number of sources to validate that localization is accurate even for multiple concurrently active sources. In each case, simulations were performed over 500 Monte-Carlo random samples and source time courses were modeled with 50 time points sampled as mixtures of sinusoidal signals.

\begin{figure*}[ht!]
    \centering
    \captionsetup{position=top}
    \subfloat[Three sources with inter-source correlation 0.5]{\includegraphics[trim=0.5cm 0.0cm 0.75cm 0.0cm, clip,scale=0.55]{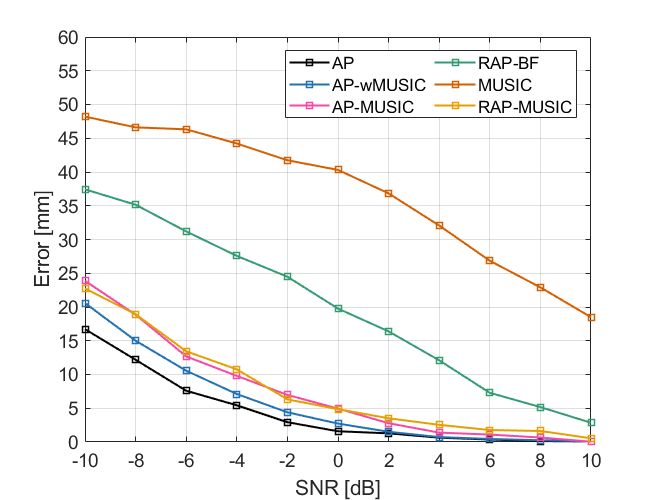}}
    \subfloat[Five sources with inter-source correlation 0.5]{\includegraphics[trim=0.5cm 0cm 0.75cm 0cm, clip,scale=0.55]{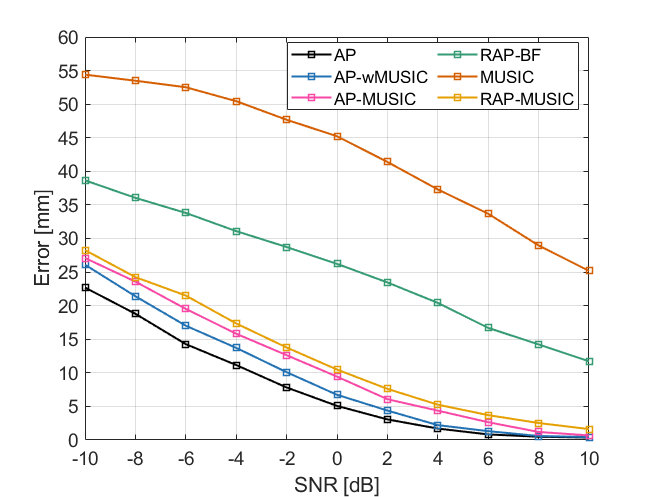}}
    \caption{Localization error for multiple simultaneously active sources. Monte Carlo simulations were conducted for the case of $Q=3$ (a) and $Q=5$ (b) sources. In all cases, pairwise inter-source correlation was set to $\rho=0.5$.}
    \label{fig:number_sources}
\end{figure*}

\subsection{Performance evaluation with a real phantom}

We assessed the performance of the AP localization algorithm using the phantom data provided in the phantom tutorial\footnote{https://neuroimage.usc.edu/brainstorm/Tutorials/PhantomElekta} of the Brainstorm software \cite{Brainstorm2011}. The phantom device was provided with the Elekta Neuromag system and contained 32 artificial dipoles in locations dispersed in four quadrants (Figure \ref{fig:geometry}b). The phantom dipoles were energized using an internal signal generator and an external multiplexer box was used to connect the signal to the individual dipoles. Only one dipole could be activated at a time, therefore we combined data trials from different dipoles to simulate the concurrent activation of multiple synchronous sources. 

The phantom data were collected using an Elekta Neuromag MEG system (306-channel probe unit with 204 planar gradiometer sensors and 102 magnetometersensors). The source space comprised the volume of a sphere, centered within the MEG sensor array with 64.5 mm radius, sampled with a regular (isotropic) grid of points with 2.5 mm resolution yielding a total of 56,762 grid points. The lead field matrix was estimated using BrainStorm \cite{Brainstorm2011} based on a single sphere head model.

\section{Results and Discussion}

\subsection{AP method has low localization error across an extended SNR range}

We assessed the performance of AP method against scanning methods for different SNR levels ranging from -10 dB to 10 dB. Monte Carlo simulation analyses were conducted with the geometry of a real adult human subject and a 306-channel MEG sensor array. Performance was evaluated for the case of two active sources with inter-source correlation values $\rho=0$ and $\rho=0.5$ (Figure \ref{fig:snr}).

The AP method had the lowest localization error in all SNR levels until performance saturated at zero localization error for high SNR values. The second best method was AP-wMUSIC, which consistently outperformed the remaining methods. It was followed by AP-MUSIC, which performed similarly to RAP-MUSIC but with better performance at very low SNR values. Collectively, all three AP-based methods fared better than MUSIC, RAP-MUSIC, and RAP-beamformer, providing evidence that the AP algorithm offers superior performance than popular scanning methods. The worse performance was obtained from the MUSIC algorithm, demonstrating its disadvantage against iterative (recursive or AP-based) multi-source methods. All results held true for both uncorrelated ($\rho=$0) and correlated ($\rho=$0.5) sources.

\subsection{AP method is robust across the entire inter-source correlation range}

We extended the above simulations to evaluate the performance of the localization methods against the entire range of inter-source correlation values from  uncorrelated ($\rho=0$) to perfectly synchronous ($\rho=1$) (Figure \ref{fig:correlation}). For the low SNR (-10 dB) condition, the AP method was consistently better than all other methods in all correlation values, excluding the perfectly synchronous case where the AP-wMUSIC had a small advantage. For high SNR (0 dB) most methods closed the performance gap, with AP-wMUSIC matching the AP method performance throughout. The results demonstrate that the AP method was reliably robust for the entire range of SNR and inter-source correlation values.

\subsection{AP method can localize multiple sources}

We further evaluated the performance of AP method in the case of more than two simultaneously active sources. We conducted Monte Carlo simulations for the case of $Q=3$ and $Q=5$ sources with strong inter-source correlation ($\rho=0.5$) (Figure \ref{fig:number_sources}). Consistent with the previous results for a pair of sources, we observed the same relative performance of all methods for the case of multiple sources. The AP method yielded the smallest localization error throughout, followed by the other two AP-based localization methods, though RAP-MUSIC was comparable to AP-MUSIC in some cases. 

\subsection{Performance evaluation for perfectly synchronous sources and known signal subspace rank}

\begin{figure*}[ht!]
    \centering
    \captionsetup{position=top}
    \subfloat[3 synchronous sources, signal subspace truncated to 3 components]{    \includegraphics[trim=0.5cm 0.0cm 0.75cm 0.0cm, clip,scale=0.55]{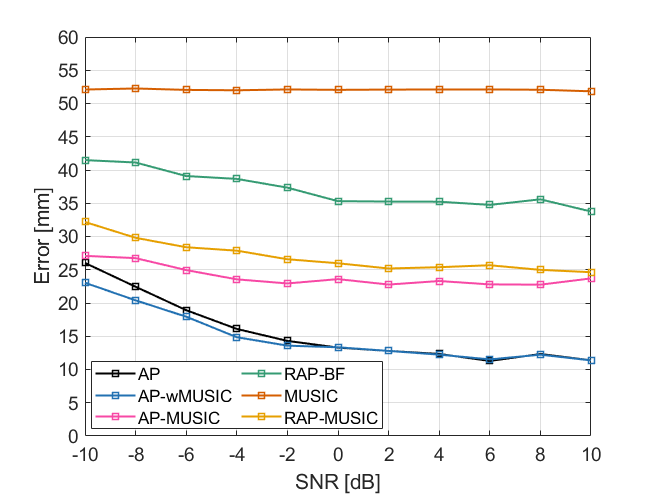}}
    \subfloat[3 synchronous sources, signal subspace truncated to 1 component]{ 
    \includegraphics[trim=0.5cm 0cm 0.75cm 0cm, clip,scale=0.55]{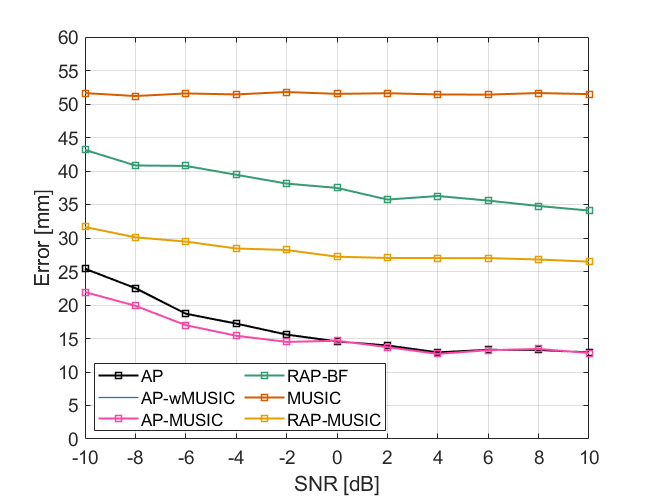}}
    \\
    \subfloat[5 synchronous sources, signal subspace truncated to 5 components]{      \includegraphics[trim=0.5cm 0cm 0.75cm 0cm, clip,scale=0.55]{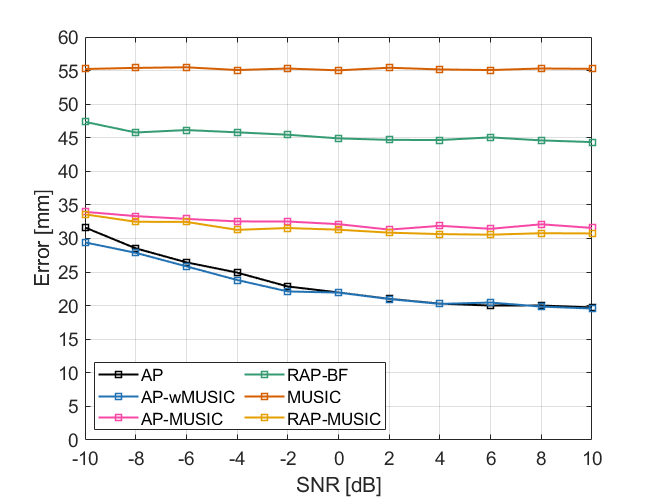}}
    \subfloat[5 synchronous sources, signal subspace truncated to 1 component]{     \includegraphics[trim=0.5cm 0cm 0.75cm 0cm, clip,scale=0.55]{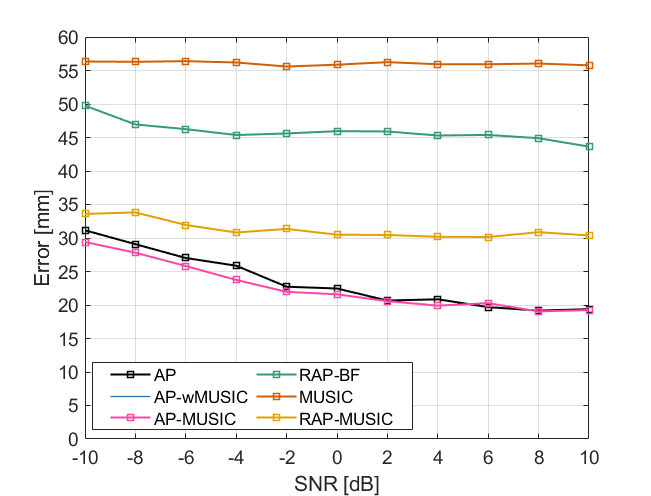}}
    \caption{Localization error vs. SNR with $Q=3$ (ab) and $Q=5$ (cd) perfectly synchronous sources ($\rho=1$). The rank of the data covariance subspace was hard truncated to a reduced number of components as indicated in each subfigure.}
    \label{fig:subspace}
\end{figure*}

The previous results revealed no localization method exceeded the performance of the AP method, with sole exception the AP-wMUSIC for perfectly synchronous sources. Here we investigated this case in more detail. We conducted Monte Carlo simulations with $Q=3$ and $Q=5$ sources having perfect inter-source correlation ($\rho=1$) (Figure \ref{fig:subspace}). To explore the effect of signal subspace, we applied hard truncation to the eigenspectrum of the data covariance matrix, leaving 3 components (three sources), 5 components (five sources), or 1 component (three or five sources). The latter matched the true rank of the signal covariance matrix, given the perfectly synchronous source time courses.

In all cases, the AP method had nearly identical localization error to AP-wMUSIC for high SNR values ($>0$ dB) (Figure \ref{fig:subspace}a-d). However, its performance dropped for low SNR values where the AP-wMUSIC emerged as the clear winner. The AP-MUSIC had inferior performance when the data covariance leaked signal from noisy eigencomponents (Figure \ref{fig:subspace}ab), but by definition matched the performance of the AP-wMUSIC when the data covariance had a single component (Figure \ref{fig:subspace}cd).

\subsection{AP method refines the localization of phantom dipoles}

We assessed the performance of the AP method in localizing real experimental data collected using the Elekta phantom. The data comprised MEG recordings collected during the sequential activation of 32 artifical dipoles. Each dipole was activated 20 times with amplitude 200 nAm and 2000 nAm, yielding a total of 20 trials in each experimental condition. The true locations of the dipoles are shown in Figure \ref{fig:geometry}b. To evaluate the ability of the AP method in localizing pairs of synchronous dipoles, we simulated the simultaneous activation of two sources by averaging the evoked responses for each of the 496 possible combinations of pairs of dipoles before source localization. AP localization relied on the freely-oriented solution (Eq. \ref{eq:p_q6}).

\begin{figure*}[ht!]
    \centering
    \captionsetup{position=top}
    \subfloat[Average AP localization error]{ 
    \includegraphics[trim=0.7cm 0cm 1cm 0cm, clip,scale=0.70]{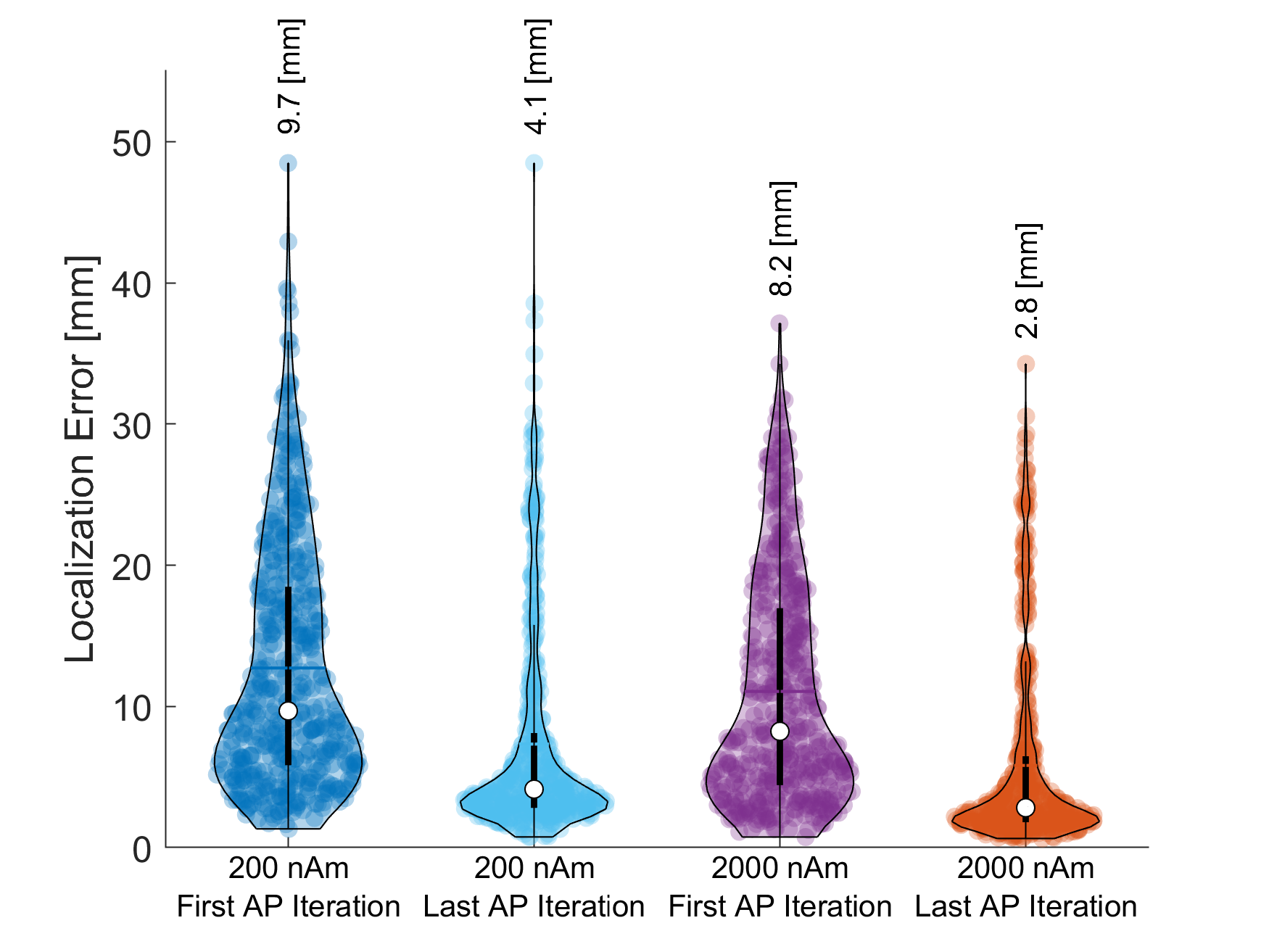}}
    \subfloat[Distribution of AP iterations]{ 
    \includegraphics[trim=0cm 0cm 1cm 0cm, clip,scale=0.6]{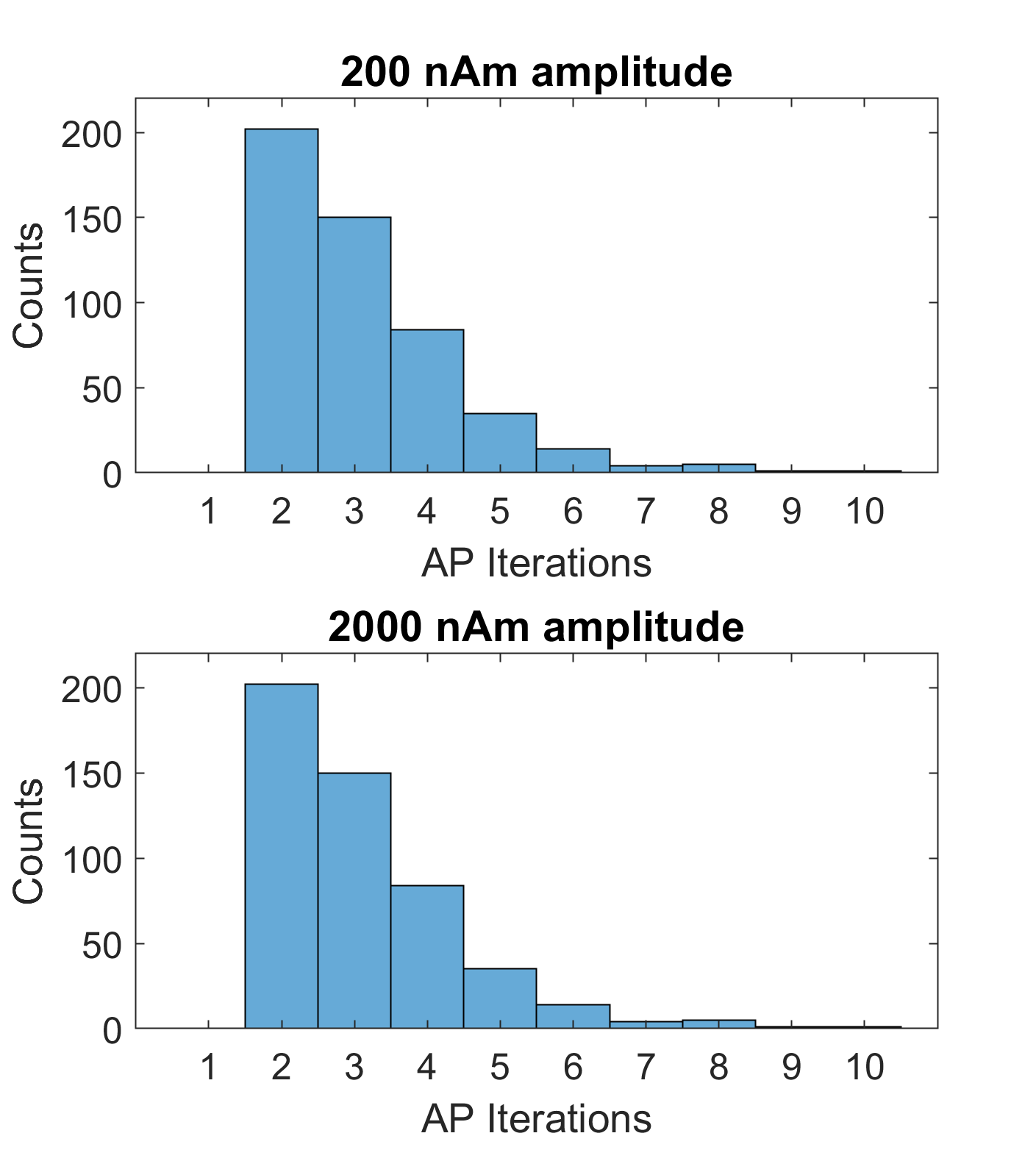}}
    \caption{Average localization error of 32 phantom dipoles. (a) Results shown for 200 nAm and 2000 nAm amplitude dipoles for the first and last iteration of the AP algorithm. (b) Histograms of AP iterations across the 496  possible combinations of pairs of dipoles.}
    \label{fig:phantom_error_all}
\end{figure*}

The average localization error across all combinations of pairs of dipoles for both the 200 nAm and 2000 nAm conditions in shown in Figure \ref{fig:phantom_error_all}. To investigate the importance of AP iterations in refining localization, we computed the error separately for the first and last iteration of the AP algorithm. We used violin plots that combine the advantages of box plots with density traces. Each "violin" contains a box-plot (white dot, vertical thick black box and thin black line), where the white dot represents the median of the distribution, the vertical thick black box indicates the inter-quartile range, and the thin black line denotes the extensions to the maximum and minimum values. The median value is also written on the top of the violin for convenience. The mean value is depicted by a horizontal line. The shaded areas surrounding the box plot show the probability density of the data across all pairs of dipoles, with individual data points plotted as colored circles. 

\begin{figure*}[ht!]
    \centering
    \captionsetup{position=top}
    \subfloat[200 nAm, First AP iteration]{    \includegraphics[trim=1.5cm 0cm 1.5cm 0cm, clip,scale=0.10]{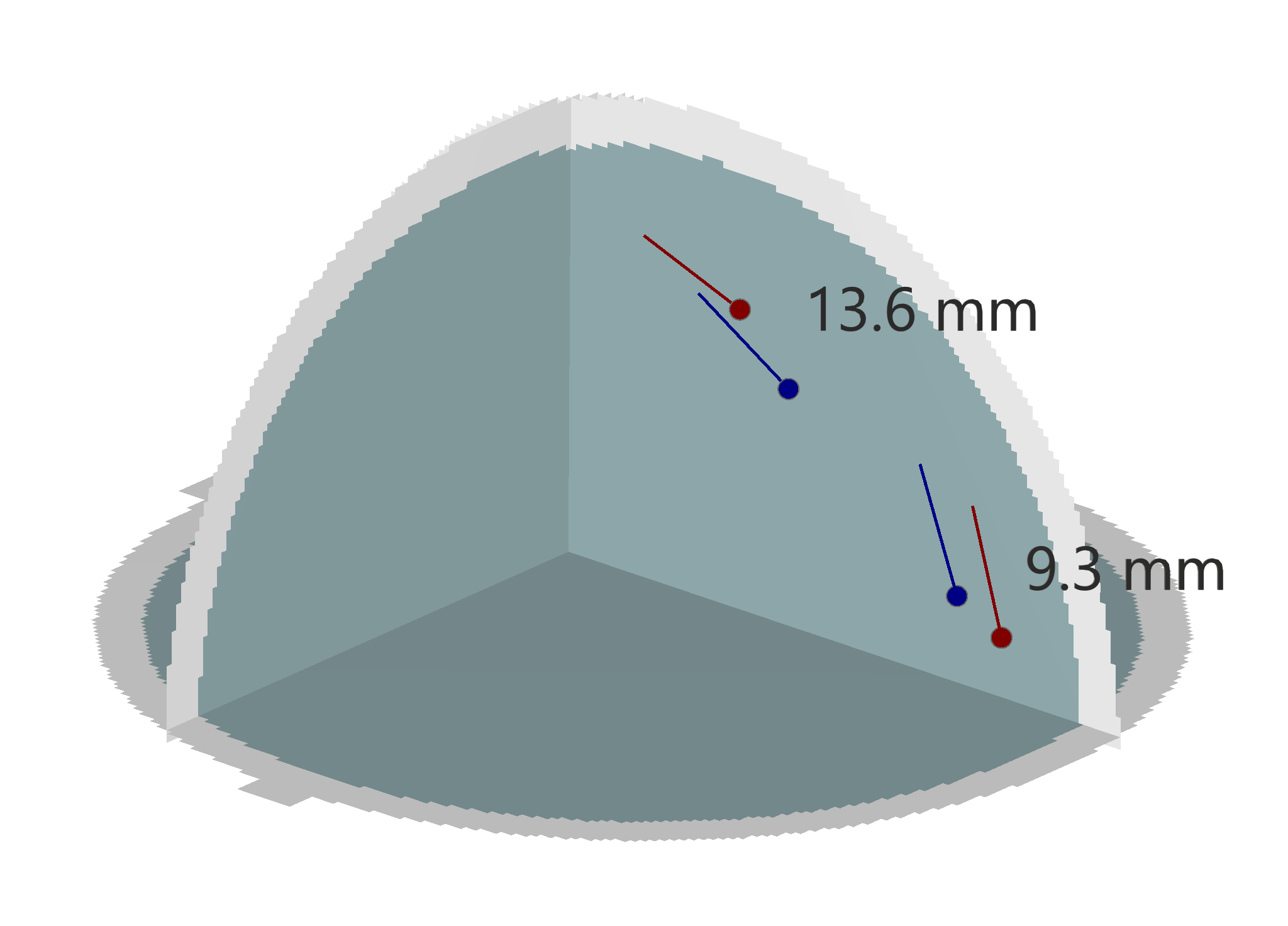}}
    \subfloat[200 nAm, Last AP iteration]{      \includegraphics[trim=-1.5cm 0cm -1.5cm 0cm, clip,scale=0.10]{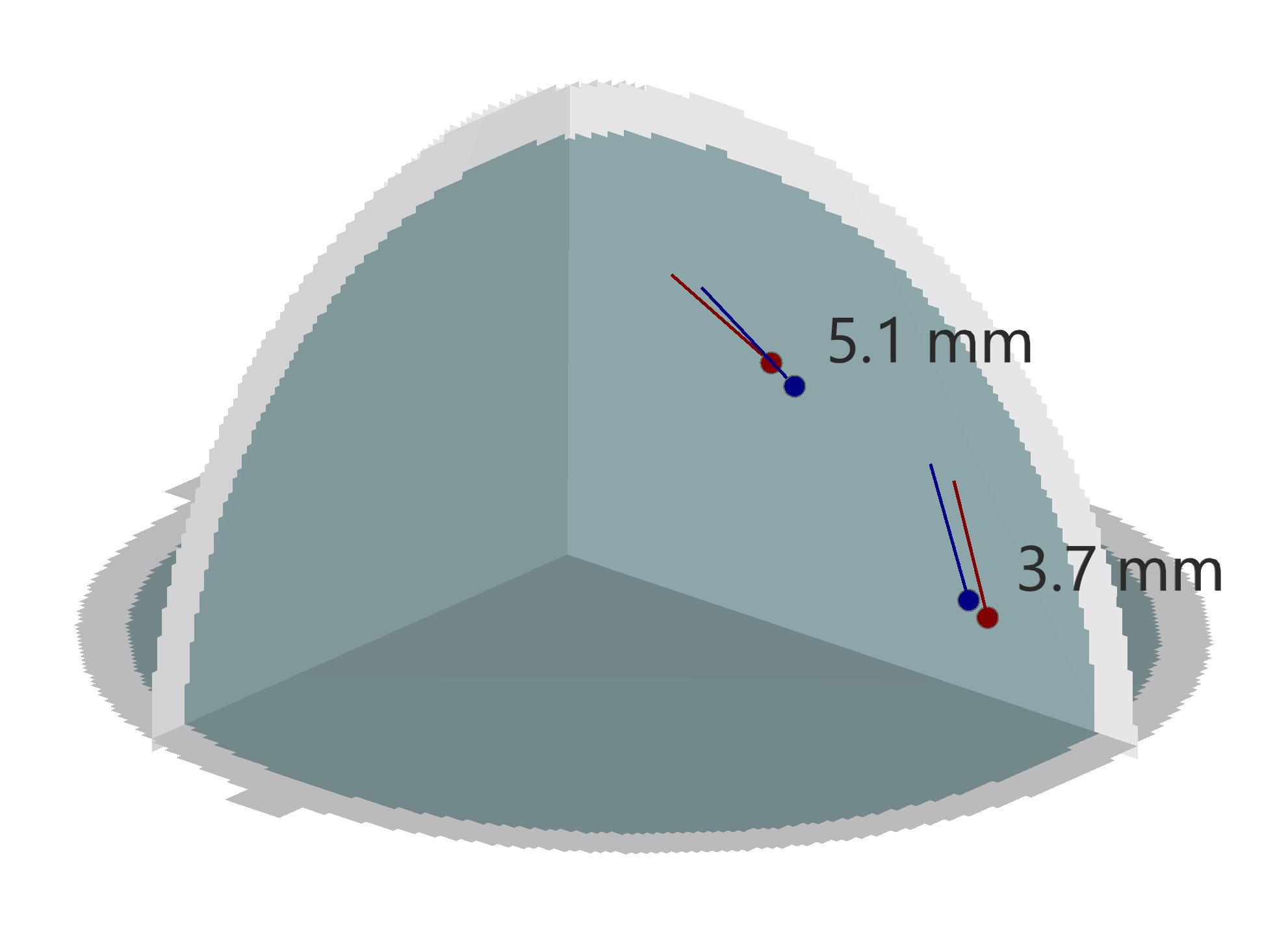}}
    \\
    \subfloat[2000 nAm, First AP iteration]{ 
    \includegraphics[trim=0.5cm 0.5cm 0.5cm 0cm, clip,scale=0.10]{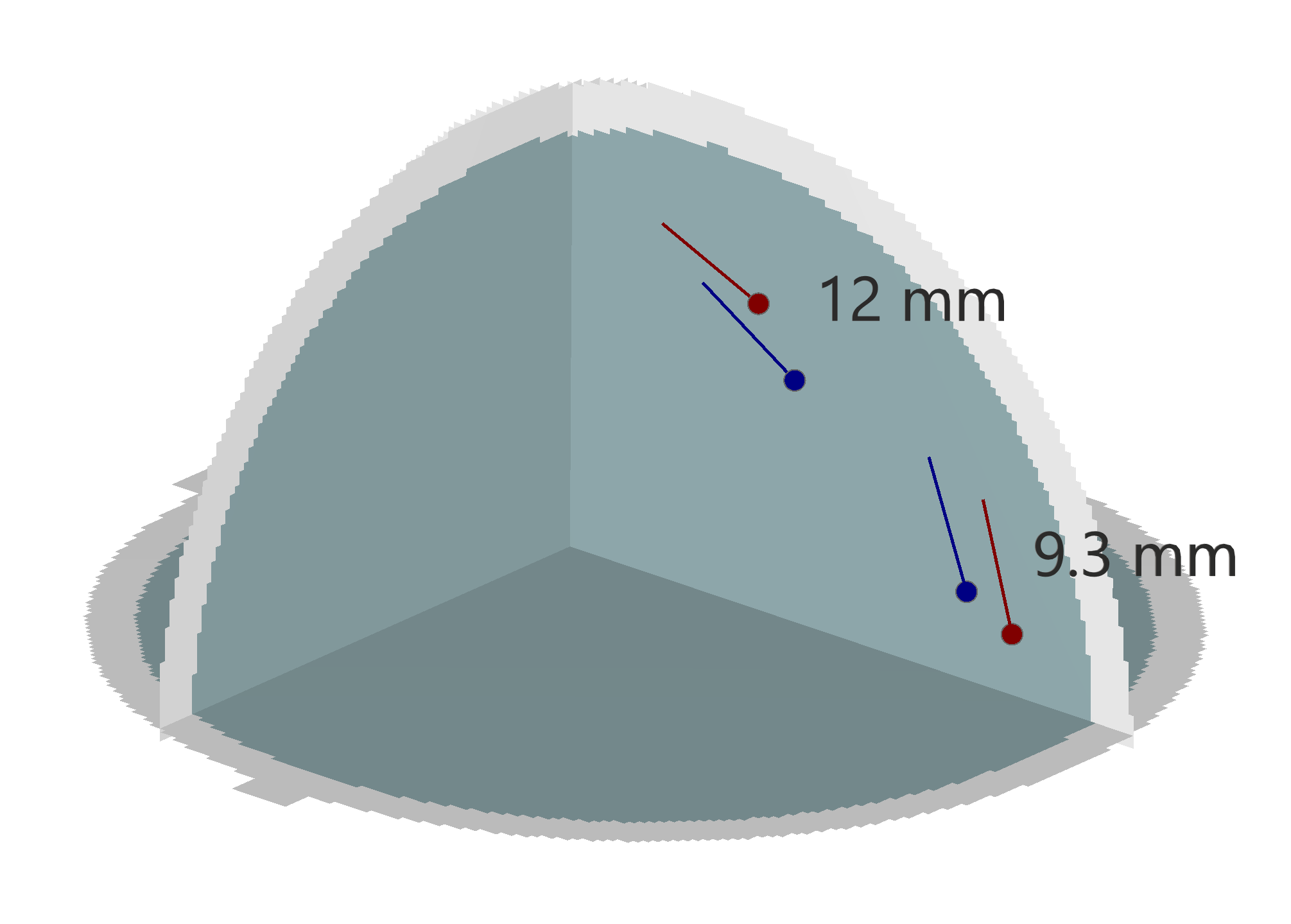}}
    \subfloat[2000 nAm, Last AP iteration]{     \includegraphics[trim=0.5cm 0.5cm 0.5cm 0cm, clip,scale=0.10]{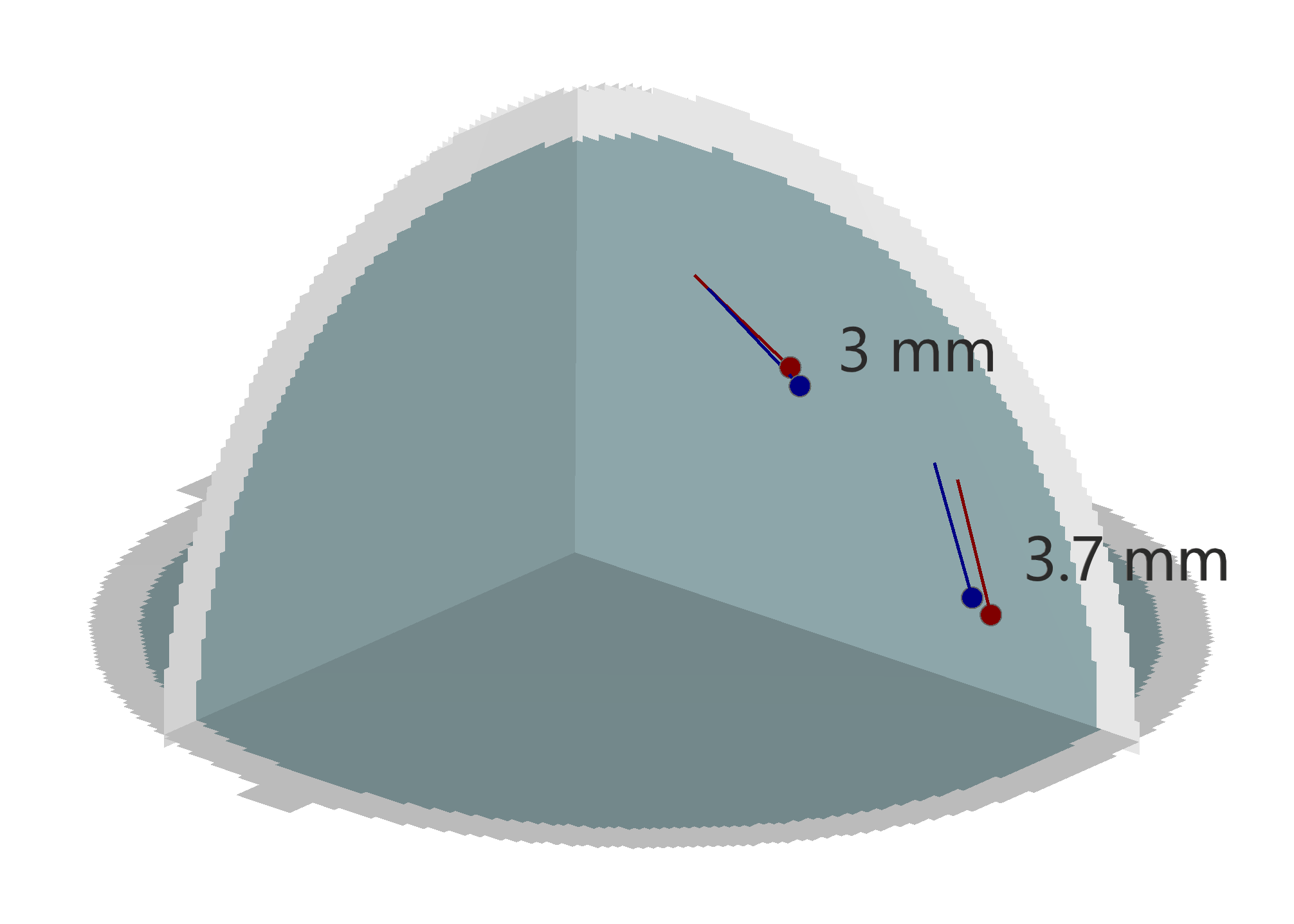}}
    \caption{Example localization of two phantom dipoles of 200 nAm or 2000 nAm amplitude for the first and last AP iteration.}
    \label{fig:example_dipoles}
\end{figure*}

Localization error dropped considerably from the first to the last iteration of the AP algorithm (Figure \ref{fig:phantom_error_all}a). For the 200 nAm condition, the median localization error dropped from 9.7 mm to 4.1 nAm, and for the 2000 nAm condition from 8.2 mm to 2.8 mm. The number of required AP iterations were similar for the 200 nAm and 2000 nAm conditions, with an average of 3.10 and 2.97 iterations respectively, indicating fast convergence of the AP algorithm (Figure \ref{fig:phantom_error_all}b). The quality of  dipole localization is exemplified in Figure \ref{fig:example_dipoles}, showing that errors are considerably reduced from first to the last AP iteration for both the 200 nAm and 2000 nAm conditions.

\subsection{Characterizing the phantom performance of the AP method}

\begin{figure}[ht!]
    \centering
    \includegraphics[trim=0.5cm 0cm 0.5cm 0cm, clip,scale=0.6]{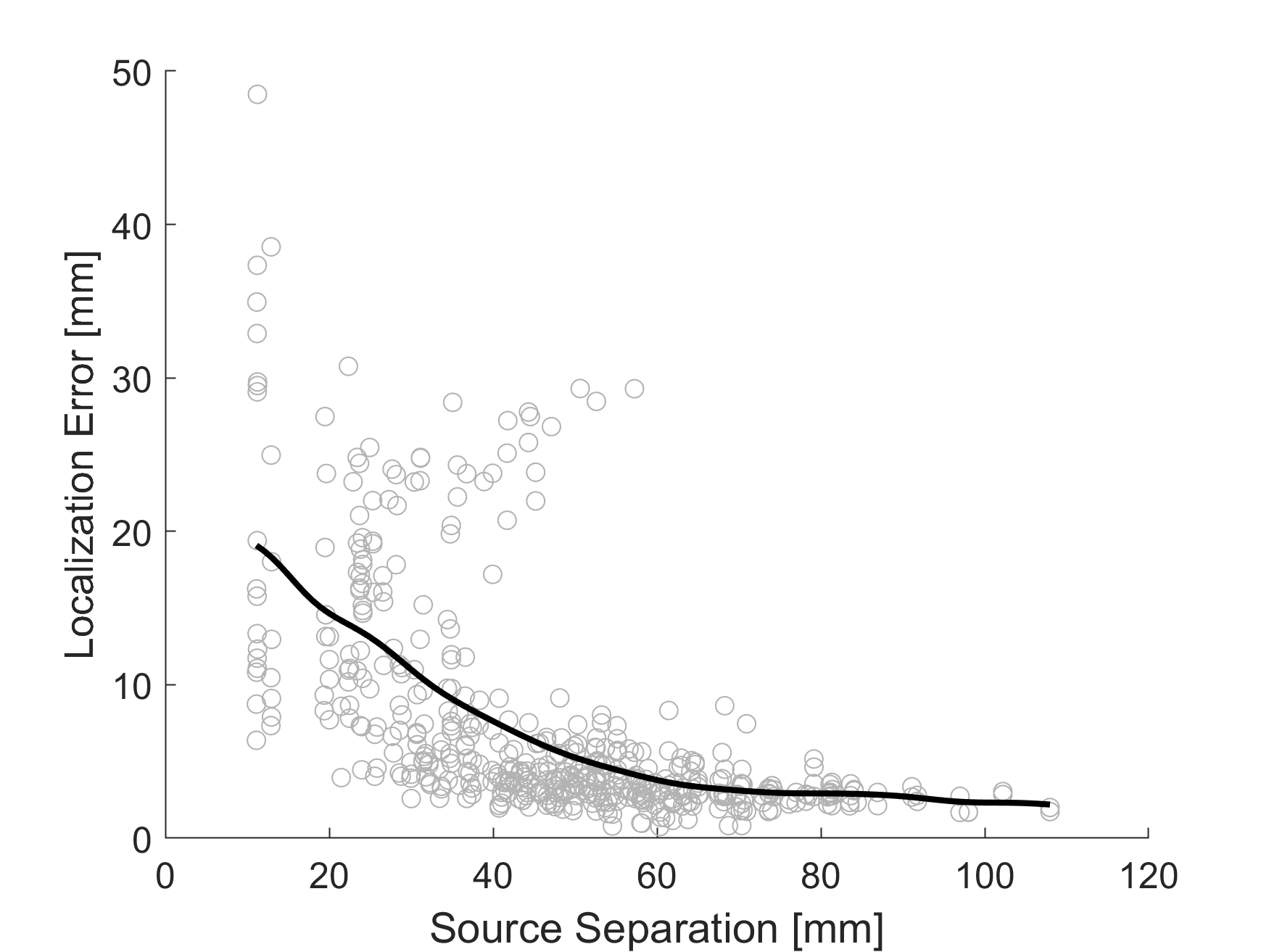}
    \caption{Localization error versus source separation for the phantom dipoles in the 200 nAm condition.}
    \label{fig:sourceseparation}
\end{figure}

We further explored the factors affecting the quality of source localization of the AP method. Localization error was higher for sources with small spatial separation, and progressively decreased as the true dipoles were further apart (Figure \ref{fig:sourceseparation}). This was expected, as nearby sources have similar topographies and are hard to dissociate with each other. 

\begin{figure*}[ht!]
    \centering
    \captionsetup{position=top}
    \subfloat[Localization error vs. required AP iterations]{    \includegraphics[trim=0.5cm 0cm 0.5cm 0cm, clip,scale=0.6]{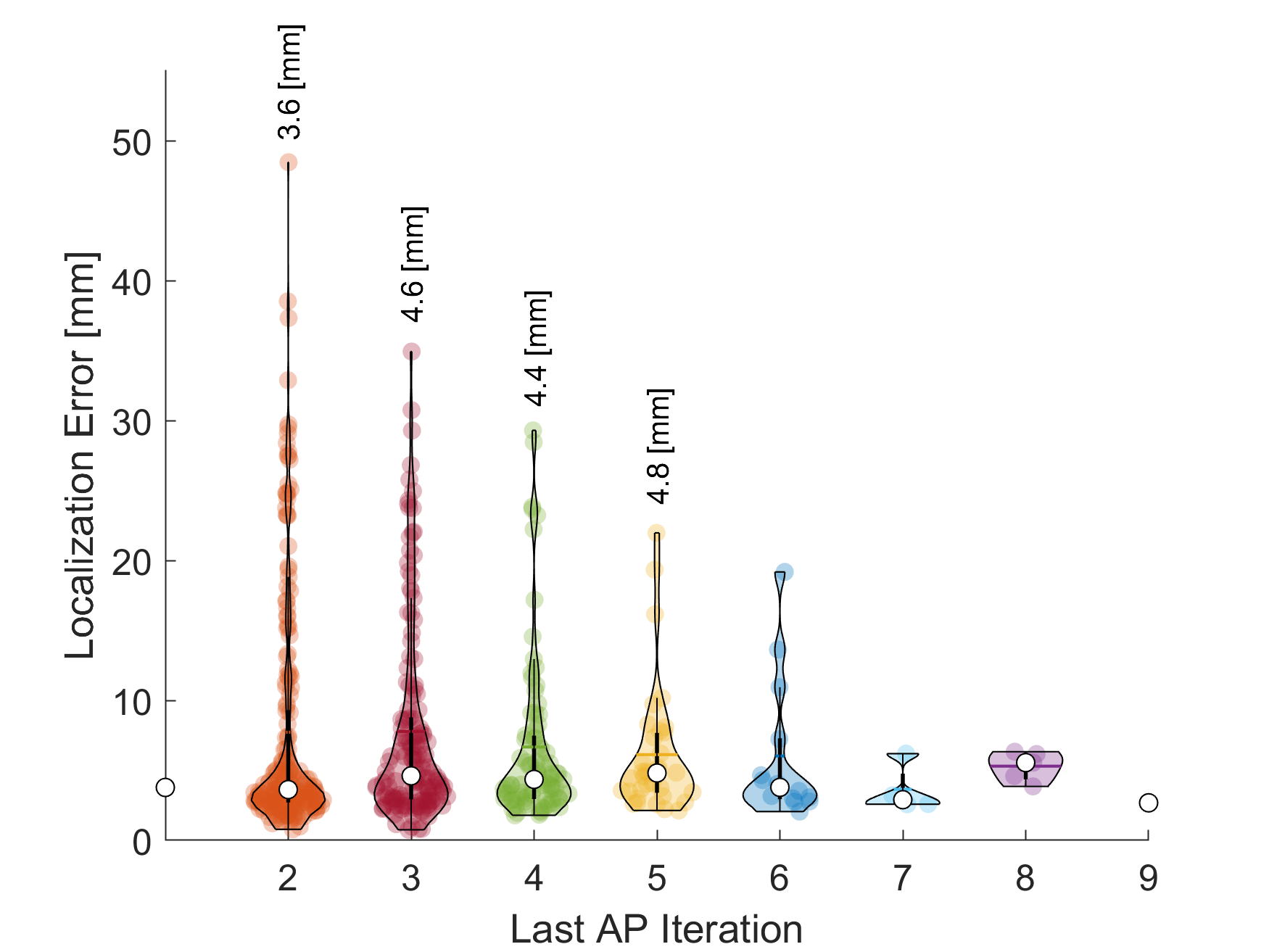}}
    \subfloat[Reduction of localization error vs. required AP iterations]{ 
    \includegraphics[trim=0.5cm 0cm 0.5cm 0cm, clip,scale=0.6]{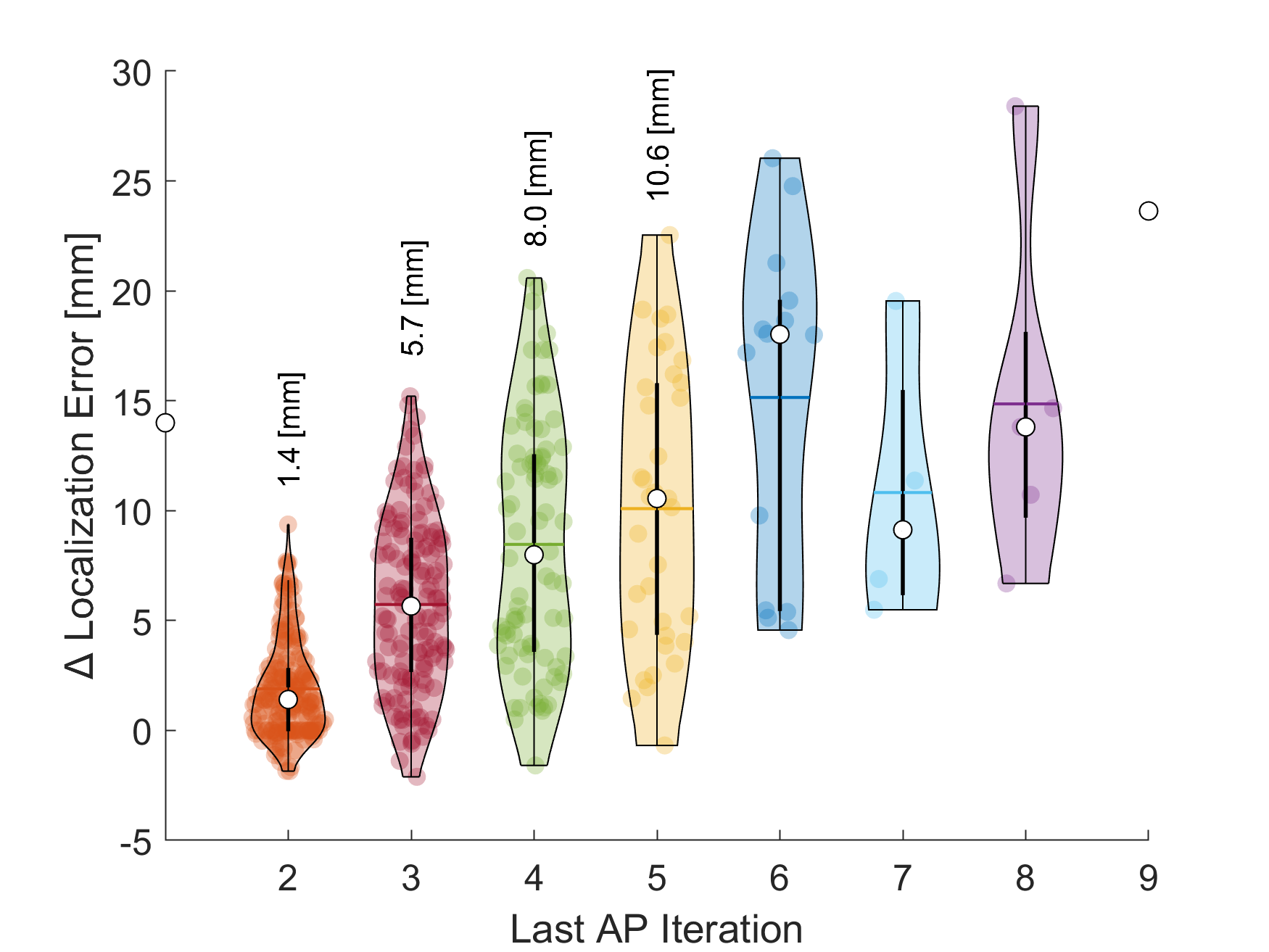}}
    \\
    \subfloat[Localization error vs. source order]{     \includegraphics[trim=0.5cm 0cm 0.5cm 0cm, clip,scale=0.6]{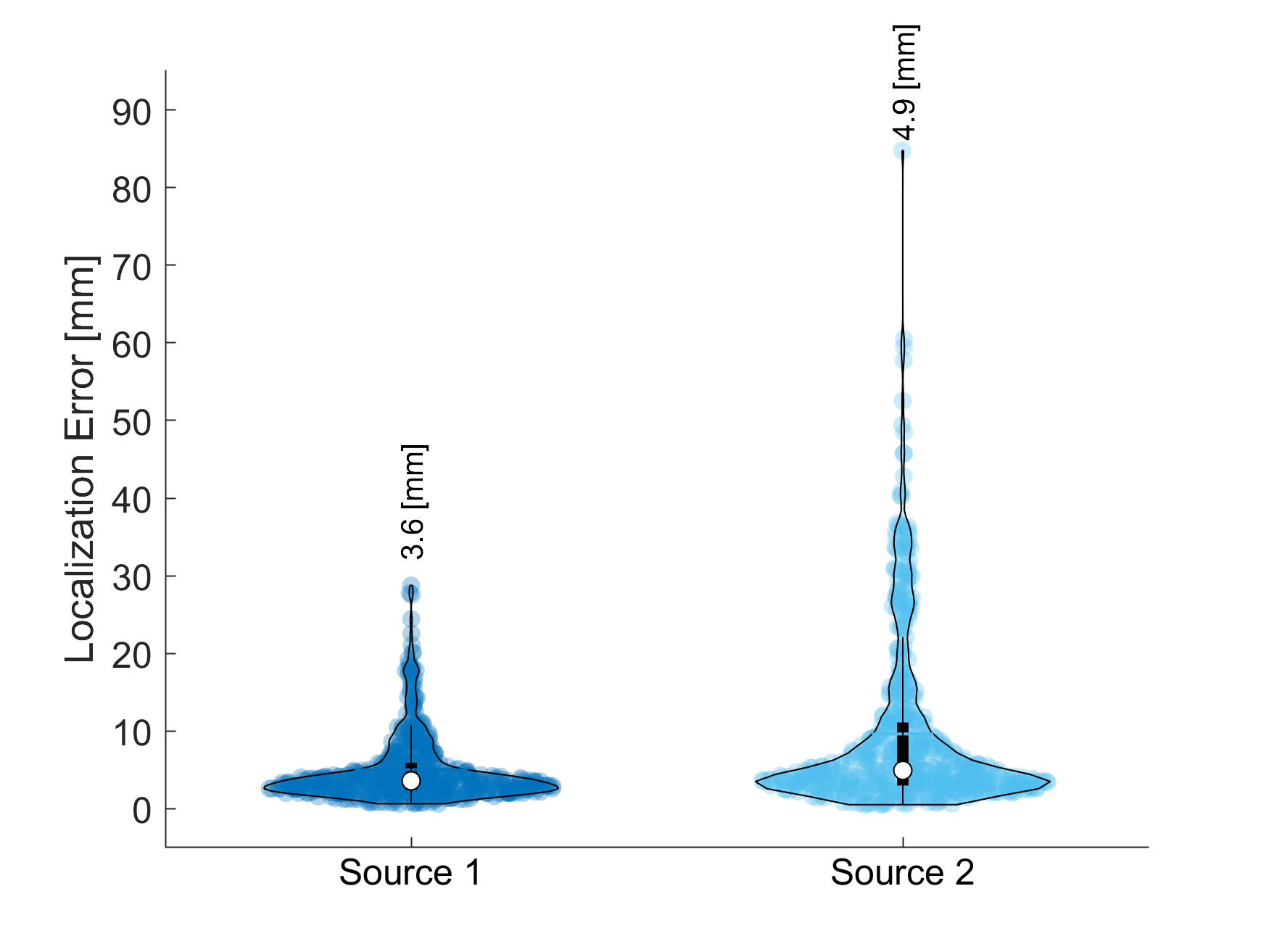}}
    \subfloat[Reduction of localization error  vs. source order]{      \includegraphics[trim=0.5cm 0cm 0.5cm 0cm, clip,scale=0.6]{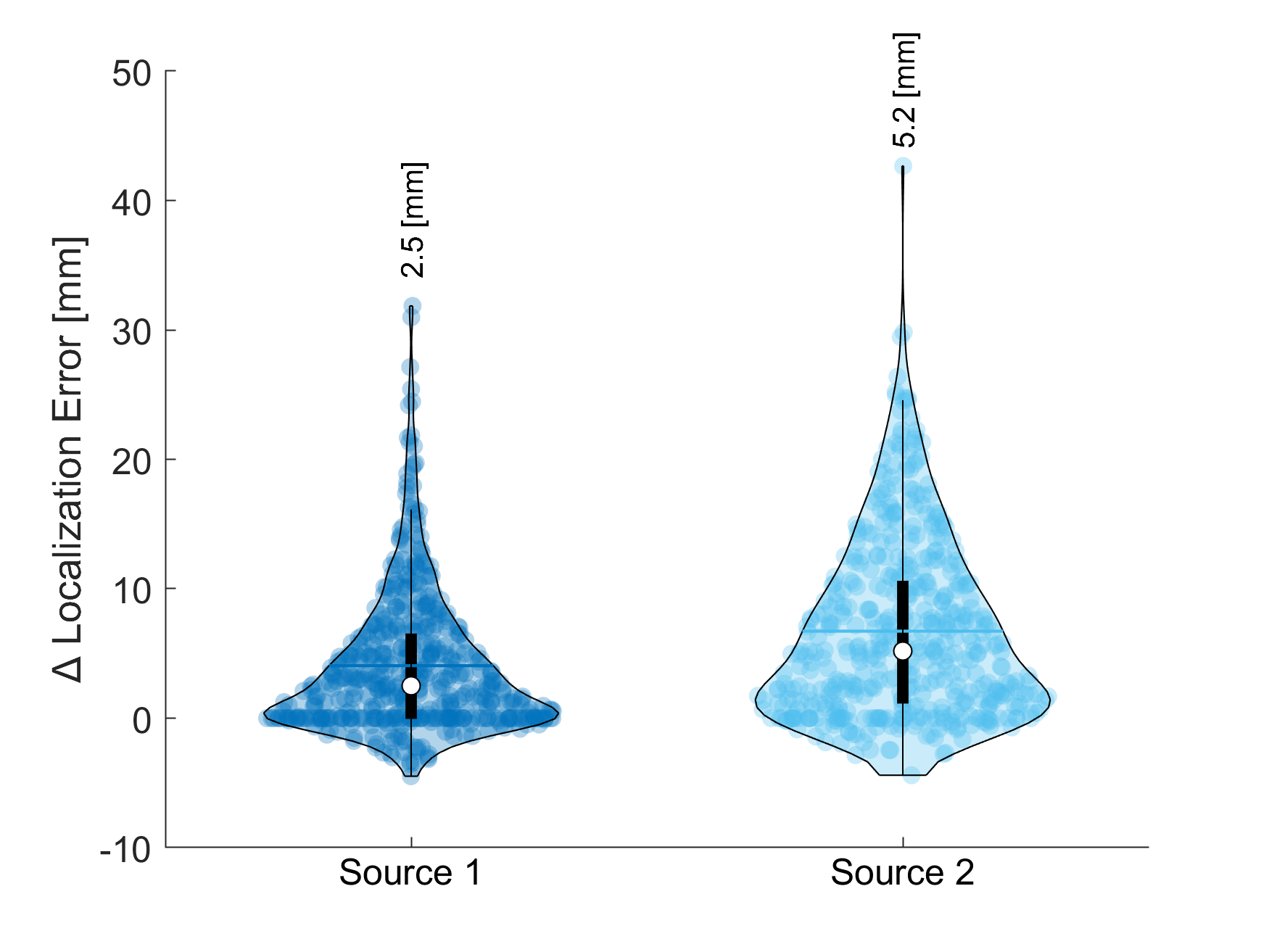}}    
    \caption{Performance of AP algorithm for the 200 nAm phantom dataset. (a) Localization error for different values of the last AP iteration. (b) Reduction of localization error for different values of the last AP iteration. (c) Localization error for the first and second source. (d) Reduction of localization error for the first and second source.}
    \label{fig:APcharacteristics}
\end{figure*}

The final localization error was not dependant on the required number of AP iterations, suggesting that the quality of source localization did not depend on the initialization of the algorithm (Figure \ref{fig:APcharacteristics}a). Consistent with this idea, the reduction of localization error ($\Delta$ Localization Error) was higher with increased number of AP iterations, indicating that each iteration progressively leads to improved localization performance (Figure \ref{fig:APcharacteristics}b). 

The final localization error was smaller for the first than the second detected source (Figure \ref{fig:APcharacteristics}c). This is consistent with the initialization of the AP method, which first detects the source that achieves the best LS fit and then progressively detects other sources. Despite this difference in localization error between the two sources, the reduction of localization error over iterations was higher for the second than the first source (Figure \ref{fig:APcharacteristics}d).

\section{Conclusion} 

We have presented a new sequential and iterative solution to the localization of MEG/EEG signals based on minimization of the LS criterion by the AP algorithm. This solution is applicable to the case of synchronous sources and even a single time sample, and is computationally similar to all the competing recursive  scanning algorithms based on the beamformer and MUSIC frameworks with the difference that the algorithm iterates till convergence. As a by-product, we have also derived signal subspace versions of this algorithm, referred to as AP-wMUSIC  and AP-MUSIC, which are applicable even in the case of synchronous sources. 

Our simulation and real phantom results demonstrated that the AP method performs robustly for a wide range of SNR values, across the entire range of inter-source correlation values, and for multiple sources. In all simulation experiments, the AP method had consistently the best performance, excluding the case of perfectly synchronous sources where the AP-wMUSIC had a slightly better accuracy. 

Taken together, our work demonstrated the high performance of the AP algorithm in localizing MEG/EEG sources. It also revealed the importance of iterating through all sources multiple times until convergence instead of recursively scanning for sources and terminating the scan after the last source is found, which is the approach largely followed by existing scanning methods. Last, it paved the way to direct AP extensions of popular scanning methods, maximizing the impact of this work.


\section{Data Availability}
The data and analysis tools used in the current study are available at  \url{https://alternatingprojection.github.io/}.

\section{Acknowledgements}
We are thankful to John Mosher for insightful comments and suggestions that greatly helped us improve this work. This work was supported by the Center for Brains, Minds and Machines at MIT, NSF STC award CCF-1231216; and by Braude College of Engineering.

%% file: main.bbl
\begin{thebibliography}{10}

\bibitem{ilmoniemi2019brain}
R.J. Ilmoniemi and J.~Sarvas,
\newblock {\em Brain Signals: Physics and Mathematics of MEG and EEG},
\newblock MIT Press, 2019.

\bibitem{mosher_multiple_1992}
J.C. Mosher, P.S. Lewis, and R.M. Leahy,
\newblock ``Multiple dipole modeling and localization from spatio-temporal
  {MEG} data,''
\newblock {\em IEEE Transactions on Biomedical Engineering}, vol. 39, no. 6,
  pp. 541--557, June 1992.

\bibitem{hamalainen_magnetoencephalographytheory_1993}
Matti Hämäläinen, Riitta Hari, Risto~J. Ilmoniemi, Jukka Knuutila, and
  Olli~V. Lounasmaa,
\newblock ``Magnetoencephalography—theory, instrumentation, and applications
  to noninvasive studies of the working human brain,''
\newblock {\em Reviews of Modern Physics}, vol. 65, no. 2, pp. 413--497, Apr.
  1993.

\bibitem{huang_multi-start_1998}
M~Huang, C.J Aine, S~Supek, E~Best, D~Ranken, and E.R Flynn,
\newblock ``Multi-start downhill simplex method for spatio-temporal source
  localization in magnetoencephalography,''
\newblock {\em Electroencephalography and Clinical Neurophysiology/Evoked
  Potentials Section}, vol. 108, no. 1, pp. 32--44, Jan. 1998.

\bibitem{uutela_global_1998}
K.~Uutela, M.~Hamalainen, and R.~Salmelin,
\newblock ``Global optimization in the localization of neuromagnetic sources,''
\newblock {\em IEEE Transactions on Biomedical Engineering}, vol. 45, no. 6,
  pp. 716--723, June 1998.

\bibitem{khosla_spatio-temporal_1997}
D.~Khosla, M.~Singh, and M.~Don,
\newblock ``Spatio-temporal {EEG} source localization using simulated
  annealing,''
\newblock {\em IEEE Transactions on Biomedical Engineering}, vol. 44, no. 11,
  pp. 1075--1091, Nov. 1997.

\bibitem{jiang_comparative_2003}
Tianzi Jiang, An~Luo, Xiaodong Li, and F.~Kruggel,
\newblock ``A {Comparative} {Study} {Of} {Global} {Optimization} {Approaches}
  {To} {Meg} {Source} {Localization},''
\newblock {\em International Journal of Computer Mathematics}, vol. 80, no. 3,
  pp. 305--324, Mar. 2003.

\bibitem{darvas_mapping_2004}
F.~Darvas, D.~Pantazis, E.~Kucukaltun-Yildirim, and R.M. Leahy,
\newblock ``Mapping human brain function with {MEG} and {EEG}: methods and
  validation,''
\newblock {\em NeuroImage}, vol. 23, pp. S289--S299, Jan. 2004.

\bibitem{LCMV}
{B. D. van Veen, W. van Drongelen, M. Yuchtman, and A. Suzuki},
\newblock ``Localization of brain electrical activity via linearly constrained
  minimum variance spatial filtering,''
\newblock {\em IEEE Trans. Biomed. Eng.}, vol. 44, no. 9, pp. 867--880, Sep
  1997.

\bibitem{VerbaRobinson}
J.~Vrba and S.~E. Robinson,
\newblock ``Signal processing in magnetoencephalography,''
\newblock {\em Methods}, vol. 25, pp. 249--271, 2001.

\bibitem{CohBF}
{S. S. {Dalal} and K. {Sekihara} and S. S. {Nagarajan}},
\newblock ``Modified beamformers for coherent source region suppression,''
\newblock {\em IEEE Transactions on Biomedical Engineering}, vol. 53, no. 7,
  pp. 1357--1363, July 2006.

\bibitem{CorrBF}
{M. J. {Brookes}, C. M. {Stevenson}, G. R. {Barnes}, A. {Hillebrand}, M. I.
  {Simpson}, S. T. {Francis}, and P. G. {Morris}},
\newblock ``Beamformer reconstruction of correlated sources using a modified
  source,''
\newblock {\em NeuroImage}, vol. 34, no. 4, pp. 1454--1465, 2007.

\bibitem{NullBF}
{H. B. Hu, D. Pantazis, S. L. Bressler and R. M. Leahy},
\newblock ``Identifying true cortical interactions in {MEG} using the nulling
  beamformer,''
\newblock {\em NeuroImage}, vol. 49, no. 4, pp. 3161--3174, 2010.

\bibitem{DCBF}
{M. Diwakar and M.-X Huang and R. Srinivasan and D. L. Harrington and A. Robb
  and A. Angeles and L. Muzzatti and R. Pakdaman and T. Song, R. J. Theilmann
  and R. R. Lee },
\newblock ``Dual-core beamformer for obtaining highly correlated neuronal
  networks in meg,''
\newblock {\em NeuroImage}, vol. 54, no. 1, pp. 253--263, 2011.

\bibitem{EDCBF}
M.~Diwakar, O.~Tal, T.~Liu, D.~Harringtona, R.~Srinivasan, L.~Muzzatti,
  T.~Song, R.~Theilmann, R.~Lee, and M.-X. Huang,
\newblock ``Accurate reconstruction of temporal correlation for neuronal
  sources using the enhanced dual-core meg beamformer,''
\newblock {\em NeuroImage}, vol. 56, pp. 1918--1928, 2011.

\bibitem{MultiLCMV}
{A. Moiseev, J. M. Gaspar, J. A. Schneider and A. T. Herdman},
\newblock ``Application of multi-source minimum variance beamformers for
  reconstruction of correlated neural activity,''
\newblock {\em NeuroImage}, vol. 58, no. 2, pp. 481--496, Sep 2011.

\bibitem{MultiBeamformers}
A.~{Moiseev} and A.~T. {Herdman},
\newblock ``Multi-core beamformers: Derivation, limitations and improvements,''
\newblock {\em NeuroImage}, vol. 71, pp. 135--146, 2013.

\bibitem{POP-MUSIC}
{Hesheng Liu} and P.~H. {Schimpf},
\newblock ``Efficient localization of synchronous eeg source activities using a
  modified rap-music algorithm,''
\newblock {\em IEEE Transactions on Biomedical Engineering}, vol. 53, no. 4,
  pp. 652--661, April 2006.

\bibitem{WedgeMUSIC}
{A. Ewald, F. S. Avarvand and G. Nolte},
\newblock ``Wedge {MUSIC}: A novel approach to examine experimental differences
  of brain source connectivity patterns from {EEG/MEG} data,''
\newblock {\em NeuroImage}, vol. 101, pp. 610--624, 2014.

\bibitem{hui_identifying_2010}
Hua~Brian Hui, Dimitrios Pantazis, Steven~L. Bressler, and Richard~M. Leahy,
\newblock ``Identifying true cortical interactions in {MEG} using the nulling
  beamformer,''
\newblock {\em NeuroImage}, vol. 49, no. 4, pp. 3161--3174, Feb. 2010.

\bibitem{Mosher1999}
J.~C. {Mosher} and R.~M. {Leahy},
\newblock ``Source localization using recursively applied and projected {(RAP)
  MUSIC},''
\newblock {\em IEEE Transactions on Signal Processing}, vol. 47, no. 2, pp.
  332--340, Feb 1999.

\bibitem{makela_truncated_2018}
Niko Mäkelä, Matti Stenroos, Jukka Sarvas, and Risto~J. Ilmoniemi,
\newblock ``Truncated {RAP}-{MUSIC} ({TRAP}-{MUSIC}) for {MEG} and {EEG} source
  localization,''
\newblock {\em NeuroImage}, vol. 167, pp. 73--83, Feb. 2018.

\bibitem{makela_locating_2017}
Niko Mäkelä, Matti Stenroos, Jukka Sarvas, and Risto~J. Ilmoniemi,
\newblock ``Locating highly correlated sources from {MEG} with (recursive)
  ({R}){DS}-{MUSIC},''
\newblock preprint, Neuroscience, Dec. 2017.

\bibitem{SPMBrainMap}
S.~{Baillet}, J.~C. {Mosher}, and R.~M. {Leahy},
\newblock ``Electromagnetic brain mapping,''
\newblock {\em IEEE Signal Processing Magazine}, vol. 18, no. 6, pp. 14--30,
  Nov 2001.

\bibitem{AP}
I.~{Ziskind} and M.~{Wax},
\newblock ``Maximum likelihood localization of multiple sources by alternating
  projection,''
\newblock {\em IEEE Transactions on Acoustics, Speech, and Signal Processing},
  vol. 36, no. 10, pp. 1553--1560, Oct 1988.

\bibitem{S-MUSIC}
{S. K Oh} and {C. K. Un},
\newblock ``A sequential estimation approach for performance improvement of
  eigenstructure-based methods in array processing,''
\newblock {\em IEEE Transactions on Signal Processing}, vol. 41, no. 1, pp.
  457--463, Jan 1993.

\bibitem{MultiLCMV2}
{A. T. Herdman, A. Moiseev and U. Ribary},
\newblock ``Localizing event-related potentials using multi-source minimum
  variance beamformers: A validation study,''
\newblock {\em Brain Tomography}, vol. 31, no. 4, pp. 546--565, July 2018.

\bibitem{WeightedMUSIC}
D.~H. {Johnson},
\newblock ``The application of spectral estimation methods to bearing
  estimation problems,''
\newblock {\em Proceedings of the IEEE}, vol. 70, no. 9, pp. 1018--1028, Sep.
  1982.

\bibitem{SelfConsistent}
{F. Shahbazi, A. {Ewald}} and G.~{Nolte},
\newblock ``Self-consistent {MUSIC}: An approach to the localization of true
  brain interactions from {EEG/MEG} data,''
\newblock {\em NeuroImage}, vol. 112, May 2015.

\bibitem{31267}
M.~{Wax} and I.~{Ziskind},
\newblock ``Detection of the number of coherent signals by the {MDL}
  principle,''
\newblock {\em IEEE Transactions on Acoustics, Speech, and Signal Processing},
  vol. 37, no. 8, pp. 1190--1196, 1989.

\bibitem{fischl_2004}
Bruce Fischl, David~H. Salat, André~J.W. van~der Kouwe, Nikos Makris, Florent
  Ségonne, Brian~T. Quinn, and Anders~M. Dale,
\newblock ``Sequence-independent segmentation of magnetic resonance images,''
\newblock {\em NeuroImage}, vol. 23, pp. S69--S84, Jan. 2004.

\bibitem{Brainstorm2011}
François Tadel, Sylvain Baillet, John~C. Mosher, Dimitrios Pantazis, and
  Richard~M. Leahy,
\newblock ``Brainstorm: a user-friendly application for {MEG/EEG} analysis,''
\newblock {\em Computational intelligence and neuroscience}, 2011.

\bibitem{Huang1999}
M.~X. Huang, J.~C. Mosher, and R.~M. Leahy,
\newblock ``A sensor-weighted overlapping-sphere head model and exhaustive head
  model comparison for {MEG},''
\newblock {\em Physics in medicine and biology}, vol. 44, pp. 423–440, 1999.

\bibitem{sekihara2001}
K.~{Sekihara}, S.~S. {Nagarajan}, D.~{Poeppel}, A.~{Marantz}, and
  Y.~{Miyashita},
\newblock ``Reconstructing spatio-temporal activities of neural sources using
  an meg vector beamformer technique,''
\newblock {\em IEEE Transactions on Biomedical Engineering}, vol. 48, no. 7,
  pp. 760--771, 2001.

\end{thebibliography}
